\documentclass[10pt, aps, amsmath, twocolumn, superscriptaddress, showpacs, prb, floatfix]{revtex4-2}
\usepackage{feynmp}
\usepackage{amsmath}
\usepackage{graphicx}
\usepackage{multirow}
\usepackage{latexsym}
\usepackage{textcomp}
\usepackage{verbatim}
\usepackage{color}
\usepackage{bm}
\usepackage{subfigure}
\usepackage{everysel}
\usepackage{keyval}
\usepackage{ragged2e}
\usepackage{dsfont}
\usepackage{amssymb}
\usepackage{enumitem}
\usepackage{pstricks}
\usepackage{mathtools}
\usepackage{braket}
\usepackage{slashed} 
\usepackage{hyperref}
\hypersetup{
     colorlinks=true,
     allcolors = blue
     }

\usepackage{graphicx}
\usepackage{xcolor}
\usepackage{amsmath}
\usepackage{bbold}
\usepackage{amsfonts}
\usepackage{bbm}
\usepackage{comment}
\usepackage{orcidlink}

\newcommand{\osum}{{%
    \setbox0\hbox{\circ}%
    \rlap{\hbox to \wd0{\hss\sum\hss}}\box0
}}

\begin{document}

\title{Enhanced Transverse Electron Transport via Disordered Composite Formation}

\author{Sang J. Park\,\orcidlink{0000-0003-1684-4876}}
\thanks{These authors contributed equally to this work.}
\altaffiliation[Present address: ]{National Institute for Materials Science, Tsukuba, 305-0047, Japan.}
\affiliation{Department of Mechanical Engineering, Pohang University of Science and Technology, Pohang 37673, Korea}

\author{Hojun Lee\,\orcidlink{0000-0002-7406-1936}}
\thanks{These authors contributed equally to this work.}
\affiliation{Department of Physics, Pohang University of Science and Technology, Pohang 37673, Korea}

\author{Jongjun M. Lee\,\orcidlink{0000-0002-9786-1901}}
\altaffiliation[Present address: ]{Department of Physics and Quantum Horizons Alberta, University of Alberta, Edmonton, Alberta T6G 2E1, Canada.}
\affiliation{Department of Physics, Pohang University of Science and Technology, Pohang 37673, Korea}

\author{Jangwoo Ha\,\orcidlink{0009-0005-8678-3476}}
\affiliation{Department of Mechanical Engineering, Pohang University of Science and Technology, Pohang 37673, Korea}

\author{Hyun-Woo Lee\,\orcidlink{0000-0002-1648-8093}}
\email{Contact author: hwl@postech.ac.kr}
\affiliation{Department of Physics, Pohang University of Science and Technology, Pohang 37673, Korea}

\author{Hyungyu Jin\,\orcidlink{0000-0002-0187-0070}}
\email{Contact author: hgjin@postech.ac.kr}
\affiliation{Department of Mechanical Engineering, Pohang University of Science and Technology, Pohang 37673, Korea}

\begin{abstract}
Transverse electron transport in magnetic materials such as the anomalous Hall and Nernst effects holds promise for spintronic and thermoelectric applications. Efforts to enhance transverse transport have focused on finding quantum materials with large Berry curvature, skew scattering, or side jump. Here, we report a distinct approach based on composite formation. Using both theoretical modeling and experiments, we show that disordered composites of two ferromagnetic materials can exhibit significantly stronger transverse transport than constituent materials. This enhancement originates from meandering electron pathways created by the disordered composites. We identify the condition for this enhancement, which can be broadly satisfied across diverse material systems, offering a universal and tunable strategy to engineer large transverse responses.
\end{abstract}

\maketitle

%%%%%%%%%%%%%%%%%%%%%%%%%%%%%%%%%%%%%%%%%%%%

{\it Introduction.---}Transverse transport (TT) in solids~\cite{Nagaosa10RMP,Sinova15RMP,Klitzing20NatRevPhys, Chang23RMP,Bauer12NatMater,Boona14EES}—where the input and output electron fluxes are orthogonal—enables a wide range of technological applications, including magnetic field sensors~\cite{Tumanski13PE,Nakatani24APL}, transverse thermoelectric generators~\cite{Ikhlas17NatPhys,Sakai18NatPhys,Fu18EES,Guin19AdvMat,Sakai20Nature,Boona21JAP,Zhou21NatMater,He21Joule,Uchida22Joule,Pan22NatMater}, and next-generation spintronics devices~\cite{Hirohata20JMMM,Dieny20NatElectron,Park24Matter}. TT geometry helps overcome challenges in conventional longitudinal transport, such as limitations in electrical conductivity and the Seebeck effect. For instance, TT can boost thermoelectric power output by increasing the area perpendicular to the input heat flux, removing the need for complex electrical contacts and multi-element structures common in longitudinal geometry~\cite{Boona21JAP,Uchida22Joule}. Furthermore, TT offers insights into the physical properties of solids~\cite{Kane05PRL,Fu07PRB,Zhang10PRL,Onose10Science,Katsura10PRL,Cai23Nature,Takahagi25NatPhys}, revealing the interplay among charge, spin, and thermal fluxes.

The mechanisms of TT can be classified into intrinsic and extrinsic ones. The intrinsic mechanism~\cite{Karplus54PR,Jungwirth02PRL,Xiao10RMP} utilizes the Berry curvature determined by energy eigenvalues and eigenstates of a pure material. On the other hand, extrinsic mechanisms (skew scattering~\cite{Smit58Physica,Onoda08PRB} and side-jump mechanism~\cite{Berger70PRB}) generate TT through electron scattering by disorders and phonons. 
Thus, efforts to strengthen TT have been focused on finding quantum materials with large Berry curvature or strong scattering effects. In particular, exotic quantum materials such as topological materials~\cite{Haldane04PRL,Kane05PRL,Konig07Science,Chang13Science} received great attention.

\begin{figure}[t]
\includegraphics[width=245pt]{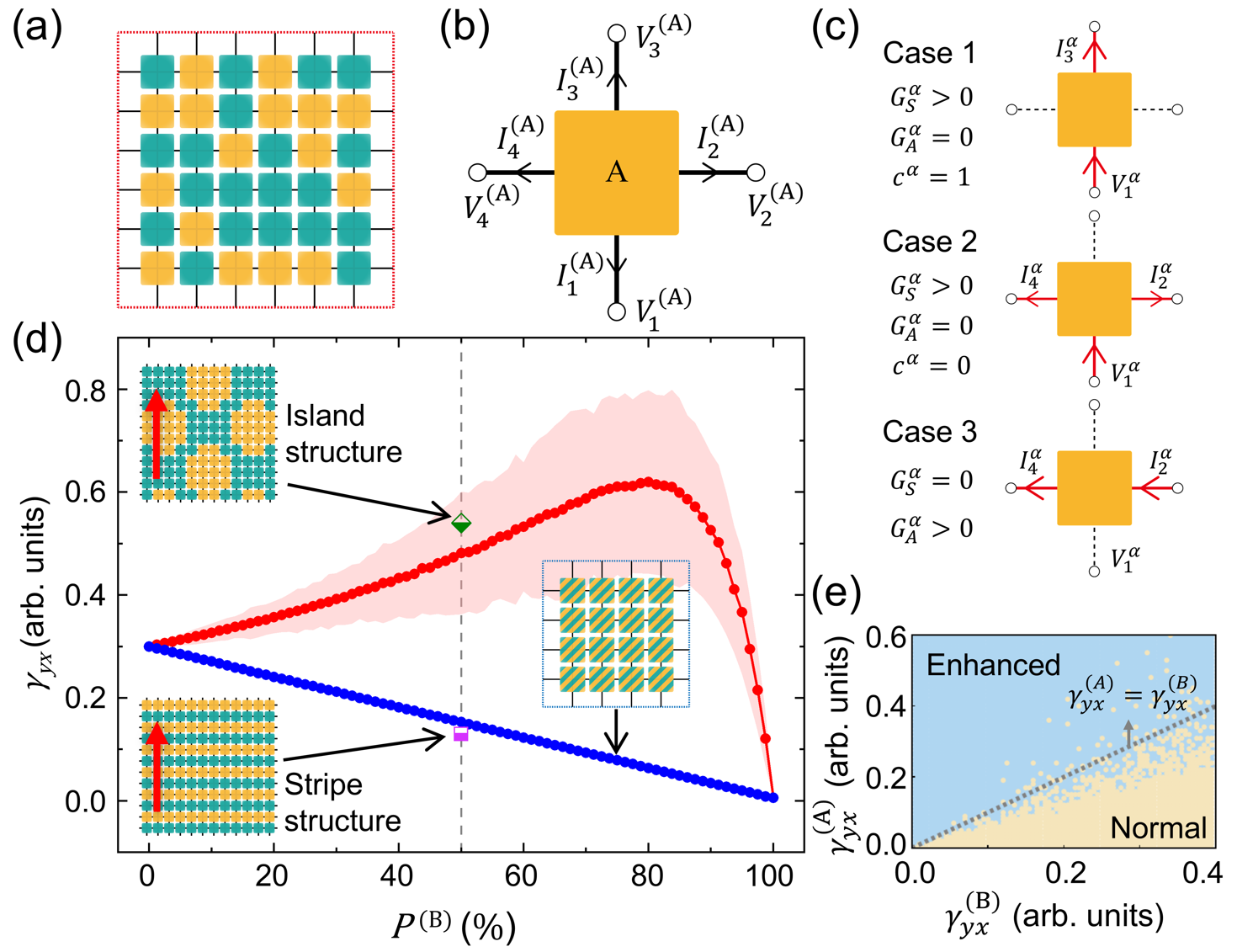}
\caption{\label{fig1}(a) Schematic of 2D network model describing the physical mixture of material A (amorphous, yellow square sites) and B (crystalline, cyan square sites). (b) Each site $\alpha$ is connected to its nearest neighbor sites through four wires $i(=1, 2, 3, 4)$. (c) Three panels illustrate current flow generated by the voltage choice, $V_1^\alpha >0$ and $V_2^\alpha=V_3^\alpha=V_4^\alpha=0$ for the three limiting cases of $(G_{\rm S}^\alpha, G_{\rm A}^\alpha, c^\alpha)$. (d) The results of the network model. Red symbols denote $\bar{\gamma}_{yx}$ and the red-shaded area represents the range of $\gamma_{yx}$ fluctuation with 1.5 times standard deviation. Blue symbols represent $\gamma_{yx}$ for the homogeneous network (right inset). Half-filled green diamond and purple square represent $\gamma_{yx}$ at $P^{(\text{B})}$= 50\% for the island and stripe networks (upper-left and lower-left insets), respectively. (e) $\gamma_{yx}^{(\text{A})}$-$\gamma_{yx}^{(\text{B})}$ diagram showing the enhancement condition. Each combination ($\gamma_{yx}^{(\text{A})}$, $\gamma_{yx}^{(\text{B})})$ is marked sky-blue if the enhancement occurs and beige otherwise. Here, $\gamma_{xx}^{({\rm A})} < \gamma_{xx}^{({\rm B})}$ is assumed.
}
\end{figure}

Our approach fundamentally differs from the previous ones. We enhance TT via composite formation---a \textit{physical} mixture of two materials [Fig.~\ref{fig1}(a)]. The composite should be distinguished from {\it chemical} compounds with intermediate chemical composition between two materials. Instead, it consists of multi-domain structures in which each domain is made of one of the two constituent materials and sufficiently large to exhibit its own longitudinal and transverse transport coefficients. Whereas the effective medium theory~\cite{Bergman91JAP} suggests that TT of a composite should lie between those of its constituents, we show theoretically that it can exceed them. The enhancement is particularly pronounced when the material with stronger TT exhibits a weaker longitudinal response. Once this condition is met, the enhancement applies broadly to various material combinations. We emphasize that the mechanism of our approach differs from the aforementioned extrinsic mechanisms in that our approach does not modify the properties of individual materials. We also demonstrate experimentally that this approach markedly enhances both the anomalous Hall effect (AHE) and the anomalous Nernst effect (ANE) of ordinary materials to a level comparable to topological single-crystalline materials. 

%%%%%%%%%%%%%%%%<Figure>%%%%%%%%%%%%%%%%
%%%%%%%%%%%%%%%%%%%%%%%%%%%%%%%%%%%%%%%%

{\it Theoretical model.---} Network models~\cite{Kirkpatrick73RMP,Stauffer79PhysRep,Monetti93PBC,Chalker88JPC,Cheianov07PRL,Albert15PRL,Rosch21ATS,Ketter25NatCommun} have been frequently used to examine transport in disordered systems. To illustrate our approach, we introduce a two-dimensional square network of sites [Fig.~\ref{fig1}(a)], where each site is connected to four nearest neighbor sites through wires. Each site represents either material A or B, and is assumed to be larger than all characteristic lengths of quantum origin. So, to describe transport in the network, we use the classical Ohm's law $I_i^{\alpha}=\sum_j G_{ij}^{\alpha}V_j^{\alpha}$ generalized to include both longitudinal transport and TT, where $V_i^{\alpha}$ and $I_i^{\alpha}$ denote voltages and currents at the wire $i$ $(=1,2,3,4)$ connected to a site $\alpha$ [Fig.~\ref{fig1}(b)], and the local conductance tensor $G_{ij}^{\alpha}$ is given by
\begin{align}\label{eq:1}
    G_{ij}^{\alpha} =
        & G_{\textrm{S}}^{\alpha} \left( c^{\alpha} \frac{ 3\delta_{i,j+2} + \delta_{i,j} - 1}{2}  - \frac{ 3\delta_{i,j} + \delta_{i,j+2} - 1 }{2}   \right) \notag
    \\
        & + G_{\textrm{A}}^{\alpha} \frac{ \delta_{i,j+1} - \delta_{i,j+3} }{2} .
\end{align}
Here, the subscripts $j+1$, $j+2$, $j+3$ of delta functions should be interpreted as $j+1\ {\rm mod}\ 4$, $j+2\ {\rm mod}\ 4$, $j+3\ {\rm mod}\ 4$, respectively.
$G_{\rm S}^\alpha$ and $c^\alpha (0 \le c^\alpha \le 1)$ characterize the symmetric part and $G_{\rm A}^\alpha$ the antisymmetric part of $G_{ij}^\alpha$. Their meanings are illustrated in Fig.~\ref{fig1}(c).
Together, they can describe arbitrary transport properties of isotropic materials. We set $(G_{\rm S}^\alpha,c^\alpha,G_{\rm A}^\alpha)$ equal to $(G_{\rm S}^{({\rm A})},c^{({\rm A})},G_{\rm A}^{({\rm A})})$ or $(G_{\rm S}^{({\rm B})},c^{({\rm B})},G_{\rm A}^{({\rm B})})$ depending on whether site $\alpha$ is occupied by material A or B. These parameters are related to the longitudinal conductivities $\gamma_{xx}^{({\rm A})}$ and $\gamma_{xx}^{({\rm B})}$, and the transverse conductivities $\gamma_{yx}^{({\rm A})}$ and $\gamma_{yx}^{({\rm B})}$ of the materials A and B (see End Matter). 

Once the material A or B is assigned randomly to each site with the probability $P^{({\rm A})}$ or $P^{({\rm B})}$ ($P^{({\rm A})}+P^{({\rm B})}=1$), the longitudinal ($\gamma_{xx}$) and transverse ($\gamma_{yx}$) conductivities of the network are evaluated under the boundary condition that amounts to the Hall measurement geometry, where the net current flows along the vertical direction [red arrows in the left insets in Fig.~\ref{fig1}(d)]. For each $P^{(\text{B})}$, multiple random networks of materials A and B are generated, and $\gamma_{xx}$ and $\gamma_{yx}$ are evaluated for each realization (see End Matter).

Figure~\ref{fig1}(d) shows the calculation result. The red dots represent the average $\bar{\gamma}_{yx}$ of $\gamma_{yx}$ over the realizations and the lightly-red-shaded area denotes the fluctuation range of $\gamma_{yx}$. Interestingly, $\bar{\gamma}_{yx}$ depends on $P^{(\text{B})}$ in a {\it nonmonotonic} way. In particular, $\bar{\gamma}_{yx}$ at an intermediate $P^{(\text{B})}$ (in the case of Fig.~\ref{fig1}(d), $P^{(\text{B})}=80\%$) is higher than the values of $\bar{\gamma}_{yx}$ at $P^{(\text{B})}=0\%$ and $100\%$. This shows that the mixture of two materials can exhibit a higher transverse conductivity than each pure material. To determine how generic the mixture-induced enhancement of $\bar{\gamma}_{yx}$ is, we vary $\gamma_{yx}^{(\text{A})}$ and $\gamma_{yx}^{(\text{B})}$ while keeping $\gamma_{xx}^{({\rm A})}$ and $\gamma_{xx}^{({\rm B})}$ fixed, and examine whether the enhancement occurs. The calculation result for the situation with $\gamma_{xx}^{({\rm A})}<\gamma_{xx}^{({\rm B})}$ [Fig.~\ref{fig1}(e)] indicates the enhancement when $\gamma_{yx}^{(\text{A})}>\gamma_{yx}^{(\text{B})}$. That is, the enhancement condition is quite simple: the material with a lower longitudinal conductivity needs to have a higher transverse conductivity.

%%%%%%%%%%%%%%%%<Figure>%%%%%%%%%%%%%%%%
\begin{figure}[t]
\includegraphics[width=245pt]{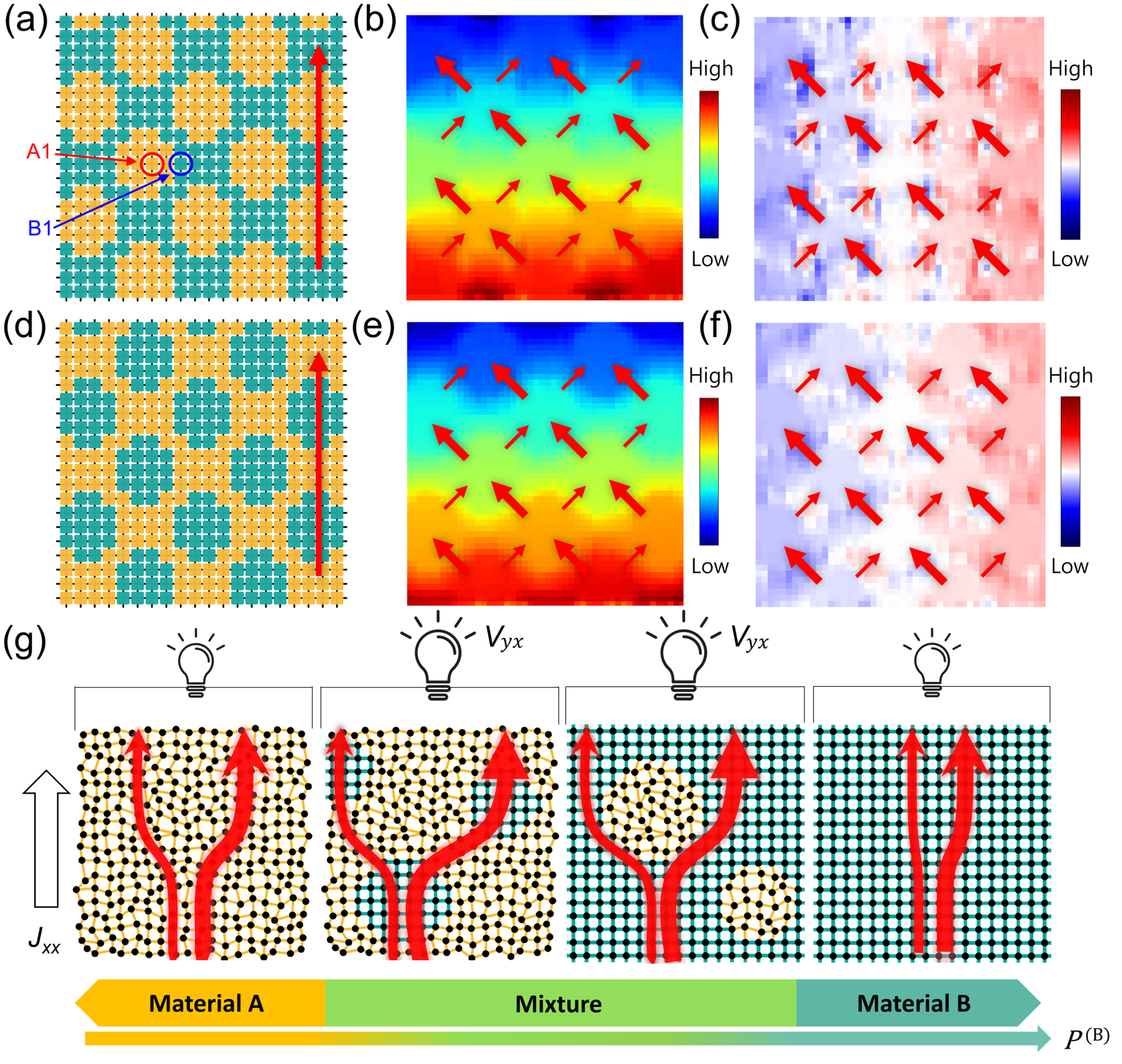}
\caption{\label{fig2}Meandering current paths originated from domain geometry. (a) Islands of material A (amorphous, yellow) embedded in material B background (crystalline, cyan).  (d) Reverse configuration. The red arrows in (a) and (d) indicate the direction of net longitudinal flux $J_{xx}$. The color in (b) and (e) shows the local voltage profiles for (a) and (d), respectively, while (c) and (f) display the transverse components of the voltage profiles (see End Matter). Red arrows in (b-c) and (e-f) denote current paths distorted by the embedded islands. (g) Schematic of TT enhancement via the composite formation.}
\end{figure}
%%%%%%%%%%%%%%%%%%%%%%%%%%%%%%%%%%%%%%%%

To track the origin of the enhancement, we consider a few specific networks and calculate their $\gamma_{yx}$’s. First, we consider a homogeneous network, where all sites are occupied by the single material C with its transport properties intermediate between A and B: $\gamma_{xx(yx)}^{(C)}=P^{(\text{A})}\gamma_{xx(yx)}^{(\text{A})}+P^{(\text{B})} \gamma_{xx(yx)}^{(\text{B})}$. That is, its $G_{ij}^{\alpha}$ is given by the $\alpha$-independent value, $P^{(\text{A})}G_{ij}^{(\text{A})} + P^{(\text{B})}G_{ij}^{(\text{B})}$. The resulting $\gamma_{yx}$ [blue dots in Fig.~\ref{fig1}(d)] interpolates $\gamma_{yx}$’s for $P^{(\text{B})}=0\%$ and $100\%$ in a monotonic way, following the rule of mixtures. This result implies that the inhomogeneity is essential for the enhancement. Next, we consider the networks with stripe structures and island structures [lower and upper left insets of Fig.~\ref{fig1}(d)], both at $P^{(\text{B})} = 50$\%. $\gamma_{yx}$ for the stripe structure network is close to $\gamma_{yx}$ for the homogeneous network, indicating that the mixture does not guarantee the enhancement. On the other hand, $\gamma_{yx}$ for the island structure network, where material A islands are embedded in material B, is almost three times larger than that of the homogeneous network and close to $\bar{\gamma}_{yx}$ for the randomly generated inhomogeneous network. To elucidate the origin of this enhancement in the island structure [Fig.~\ref{fig2}(a)], we examine its voltage and current profiles [Fig.~\ref{fig2}(b)], which indicate that the structure makes charge current paths meander: the current paths [red arrows in Figs.~\ref{fig2}(b) and (c)] tend to stay within the cyan region (material B) and form meandering paths simply because the longitudinal conductivity at each site is larger for domain B than domain A. For example, the current at site B1 [blue circle in Fig.~\ref{fig2}(a)] is found to be much stronger than that at site A1 [red circle in Fig.~\ref{fig2}(a)] (see End Matter). Thus, the current paths make ``side-jumps" to avoid material A islands. Interestingly, the left side-jumps are favored over the right side-jumps. This left-right imbalance, combined with the macroscopic size of each side-jump (comparable to the island size and much larger than atomic scale side-jumps in the conventional extrinsic side-jump mechanism), is responsible for the enhancement. We note that the imbalance is caused near the sites where straight segments of the current paths encounter islands of material A, which explains the enhancement condition $\gamma_{yx}^{(\text{A})}>\gamma_{yx}^{(\text{B})}$ [Fig.~\ref{fig1}(e)]. A similar enhancement occurs when materials A and B sites are swapped [Fig.~\ref{fig2}(d)]. Then, the current paths [red arrows in Figs.~\ref{fig2}(e) and (f)] consist of ``jumps" between neighboring islands of materials B to minimize the portion in material A. These jumps again lead to macroscopic-scale side-jumps (comparable to the distance between islands).
In the stripe structure network, in contrast, the physical mixture does not generate such side-jumps, and thus the mixture-induced enhancement is not significant there. However, in randomly generated networks, networks similar to the stripe structure network are extremely unlikely and almost all networks have meandering current paths, which indicates that the mixture-induced enhancement is quite generic if the condition $\gamma_{yx}^{(\text{A})}>\gamma_{yx}^{(\text{B})}$ is satisfied. In real materials formed by a physical mixture of materials A and B (e.g., composites), it is also highly likely that microstructures resembling the randomly generated networks are formed. Therefore, we put forward that the enhancement strategy is applicable to a wide range of materials. Figure~\ref{fig2}(g) schematically describes TT enhancement for intermediate values of $P^{({\rm B})}$.

%%%%%%%%%%%%%%%%<Figure>%%%%%%%%%%%%%%%%
\begin{figure}[t]
\includegraphics[width=245pt]{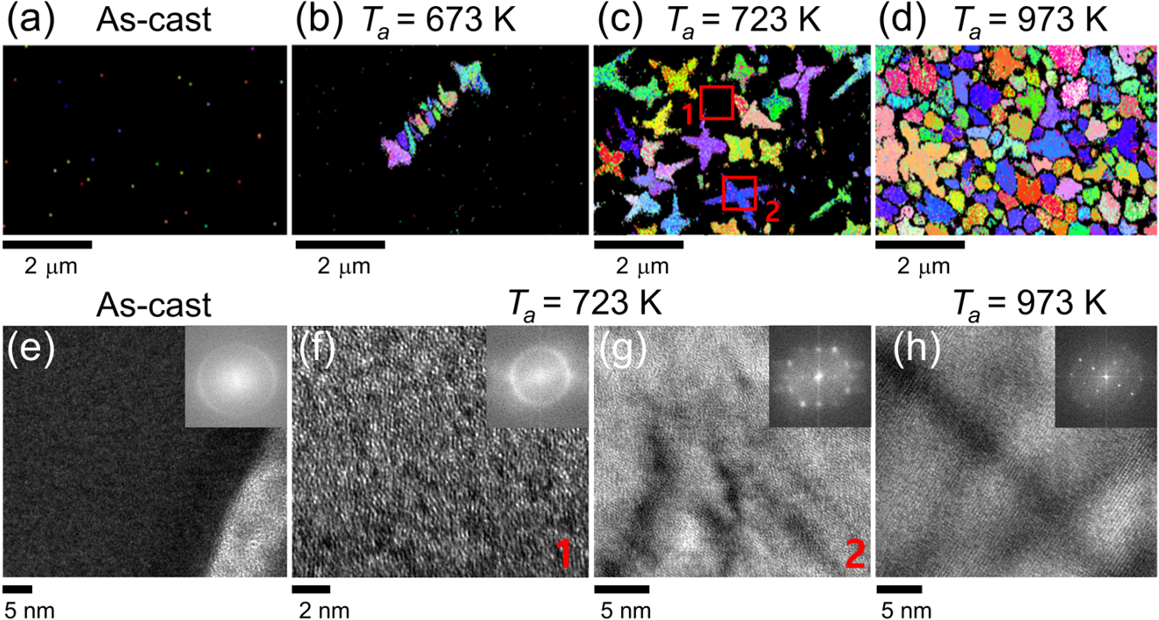}
\caption{\label{fig3}SEM color-mapping images of crystalline phases in (a) as-cast, (b) $T_{a}=673$ K, (c) $T_{a}=723$ K, and (d) $T_{a}=973$ K samples. TEM images and corresponding diffraction patterns of (e) amorphous (as-cast), (f, g) partially crystallized ($T_{a}=723$ K), and (h) poly-crystalline ($T_{a}=973$ K) states. (f) and (g) were obtained at amorphous and crystalline regions of the same $T_{a}= 723$ K sample.}
\end{figure}
%%%%%%%%%%%%%%%%%%%%%%%%%%%%%%%%%%%%%%%%

\begin{figure}[t]
\includegraphics[width=245pt]{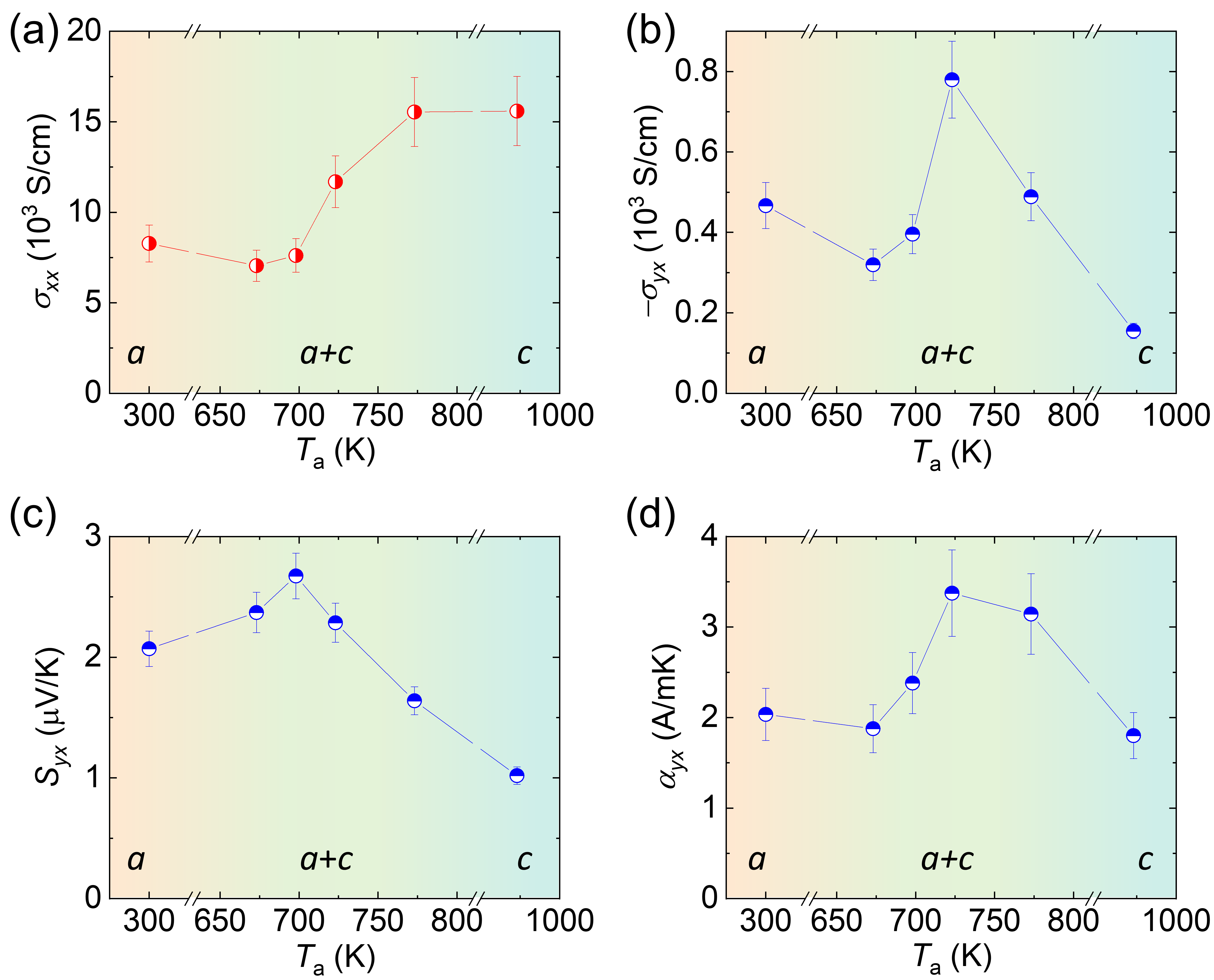}
\caption{\label{fig4}(a) Longitudinal electrical conductivity ($\sigma_{xx}$) and (b) anomalous Hall conductivity ($\sigma_{yx}$) as a function of annealing temperature ($T_{a}$). (c) Anomalous Nernst coefficient ($S_{yx}$) and (d) anomalous Nernst conductivity ($\alpha_{yx}$). All properties were measured at $T=300$ K. The symbols $a$ and $c$ denote the amorphous and crystalline phases, respectively, while $a + c$ denotes their coexistence.}
\end{figure}

{\it Experimental system: Ferromagnetic metallic glasses with controlled phase domains.---}We examined $\text{Fe}_{92.5}\text{Si}_{5}\text{B}_{2.5}$ metallic glass, a compositionally simple ferromagnet that allows direct comparison with the theoretical model. Its amorphous and crystalline phase fractions (mainly $\alpha$-Fe, Fig. S2) can be systematically tuned by varying the annealing temperature $T_{a}$ [Figs.~\ref{fig3}(a)-(d)]. The amorphous phase (as-cast, $T_a=300$ K) exhibits lower longitudinal [Fig.~\ref{fig4}(a)] and higher transverse [Fig.~\ref{fig4}(b)] conductivities than the crystalline one ($T_a=973$ K), satisfying the predicted enhancement condition. Two sets (1 h and 5 min annealing) show similar trends; the 1-h set is presented here for its clearer crystalline-amorphous domain distribution without metastable phases (5-min data in Fig. S3).

Microstructural analyses by scanning and transmission electron microscopy (SEM and TEM) reveal a gradual amorphous-to-crystalline transition with increasing $T_{a}$, from fully amorphous (as-cast) to nearly crystalline at 973 K (Fig.~\ref{fig3}). At $T_{a} = 723$ K, island-like crystalline domains emerge within an amorphous matrix, matching the mixed-domain geometry of the network model [Figs.~\ref{fig2}(g) and \ref{fig3}]. Further details on the structural characterization are provided in Figs. S2, S4, S5 and Note S2 of the Supplemental Material~\cite{suppl_ref}.

{\it Experimental validation of network model via transport measurements.---}We evaluated the TT properties to test the theoretical model. At 300 K, the amorphous sample exhibits lower longitudinal conductivity ($\sigma_{xx}=8.28 \times 10^{3}$ S/cm) and higher anomalous Hall conductivity ($\vert \sigma_{yx} \vert = 467$ S/cm) than the fully crystalline sample ($\sigma_{xx}=15.6 \times 10^{3}$ S/cm, $\vert \sigma_{yx}\vert = 155$ S/cm for $T_{a}=973$ K) [Figs.~\ref{fig4}(a) and \ref{fig4}(b)], satisfying the model condition that a component with lower $\gamma_{xx}$ and higher $\gamma_{yx}$ can yield enhancement when mixed. Here, $x$ and $y$ represent the directions of applied charge or heat current and measured transverse voltage, respectively (Fig. S6 and Ref.~\cite{Park25ATE}), following the sign convention in~\cite{Pan22NatMater}. The heterostructure samples clearly exhibit enhanced TT, with $
\sigma_{yx}$ peaking at $T_{a}=723$ K ($\vert \sigma_{yx} \vert=780$ S/cm at $T=300$ K) [Fig.~\ref{fig4}(b)], significantly exceeding both amorphous (i.e., as-cast) and crystalline (i.e., $T_{a}=973$ K) samples (see Fig. S7 for magnetic-field dependence).

The non-monotonic dependence of $\sigma_{yx}$ on $T_{a}$ contrasts with the near monotonic dependence of $\sigma_{xx}$, revealing that TT is more sensitive to the microscopic domain distribution than its longitudinal counterpart. Notably, $\vert \sigma_{yx}\vert =780$ S/cm for the $T_{a}=723$ K sample corresponds to roughly a five-fold enhancement compared with the crystalline ($T_{a}=973$ K) sample. This enhancement results in an anomalous Hall angle ($\text{AHA}=\vert \frac{\sigma_{yx}}{\sigma_{xx}}\vert$) of $6.7\%$, exceeding those of Fe-based topological single crystals such as $\text{Fe}_{3}\text{Ga}$ and $\text{Fe}_{3}\text{Al}$ ($\vert \sigma_{yx}\vert =440–540$ S/cm and $\text{AHA}=3.7–5.1\%$ at $T=300$ K)~\cite{Sakai20Nature}, demonstrating the effectiveness of our approach for designing high-performance TT materials. Note that before the enhancement, the AHA of our crystalline ($T_{a} = 973$ K) sample is merely 0.99\%.

%%%%%%%%%%%%%%%%<Figure>%%%%%%%%%%%%%%%%
%%%%%%%%%%%%%%%%%%%%%%%%%%%%%%%%%%%%%%%%

Enhanced TT is consistently observed in the thermal transport represented by the ANE at 300 K (See Fig. S8 for temperature-dependent data). The anomalous Nernst coefficient ($S_{yx}$) shows behavior similar to $\sigma_{yx}$, where the heterostructure samples outperform their amorphous and crystalline counterparts [Fig.~\ref{fig4}(c) and Fig. S3(a)]. $S_{yx}$ increases with $T_{a}$ for 1 h (5 min) annealed sample, reaching 2.67 $\mu$V/K (3.69 $\mu$V/K) for the heterostructure sample with $T_{a}=698$ K ($T_{a}=773$ K) before decreasing to 0.97 $\mu$V/K for the crystalline sample (i.e., $T_{a}=973$ K). The enhanced $S_{yx}$ values are an order of magnitude higher than those of $\alpha$-Fe single crystals ($\sim0.3$ $\mu$V/K), which is remarkable given the simple composition of the samples ($\text{Fe}_{92.5}\text{Si}_{5}\text{B}_{2.5}$). Manipulating the domain structure with only a small fraction of Si and B substitution results in such a significant enhancement in $S_{yx}$.

To further quantify the transverse thermoelectric conversion, we evaluated the anomalous Nernst conductivity  ($\alpha_{yx}=S_{xx} \sigma_{yx}+S_{yx} \sigma_{xx}$), a pivotal TT parameter that directly connects a longitudinal temperature gradient with a transverse electric current [Fig.~\ref{fig4}(d) and Fig. S3(b)]~\cite{Zhou23PRAppl,Miyasato07PRL,Xiao06PRL}. The largest value of 3.37 A/(m$\cdot$K) (4.94 A/(m$\cdot$K)) occurs for the heterostructure sample with $T_{a}=723$ K ($T_{a}=773$ K) annealed for 1 h (5 min), corresponding to approximately double (triple) that of crystalline sample (1.72 A/(m$\cdot$K) with $T_{a} = 973$ K). The observed $\alpha_{yx}$ values are comparable to, or even exceed, those of topological single crystals such as $\text{Co}_{2}\text{MnGa}$ (3–3.4 A/(m$\cdot$K))~\cite{Sakai18NatPhys,Guin19NPGAM,Xu20PRB} and $\text{Fe}_{3}\text{Ga}$ (4.49-4.83 A/(m$\cdot$K))~\cite{Sakai20Nature}, whose large $\alpha_{yx}$ values originate from intrinsic band topology.

According to our network model, TT is enhanced when straight current paths are replaced by meandering paths [second and third panels in Fig.~\ref{fig2}(g)]. In our experiment, $\sigma_{xx}$ changes rapidly near $T_{a}=723$ K, transitioning from its crystalline to the amorphous value. Therefore, meandering paths should be most pronounced near this temperature, where the experimentally observed TT enhancement agrees with the theoretical prediction. This behavior coincides with the island-like crystalline domains observed in the $T_{a}=723$ K sample [Fig.~\ref{fig3}(c)], confirming the correspondence between microstructure and macroscopic transport. 

We further investigate whether the strong enhancement of $\sigma_{yx}$, $S_{yx}$, and $\alpha_{yx}$ in our heterostructure samples could be explained by other mechanisms. We specifically consider the case of AHE which has been extensively studied over past decades~\cite{Nagaosa10RMP,Onoda08PRB,Miyasato07PRL}. When the intrinsic Berry curvature mechanism dominates AHE, $\sigma_{yx}$ should be independent of $\sigma_{xx}$~\cite{Karplus54PR}. In contrast, when the extrinsic mechanisms such as the skew scattering mechanism~\cite{Hebel64PhysLett} or the side jump mechanism~\cite{Berger70PRB} dominate, $\sigma_{yx}$ should be proportional to or independent of $\sigma_{xx}$, respectively. Comparison between experimental $\sigma_{xx}$ and $\sigma_{yx}$ [Figs.~\ref{fig4}(a) and\ref{fig4}(b)] shows that neither relation is satisfied, implying that the TT enhancement observed in our experiment does not originate from the conventional mechanisms (See detailed discussion in Notes S5 and S6, and Figs. S10-S14 in Supplemental Material~\cite{suppl_ref}). In contrast, our theoretical model naturally explains the peak formation of $\sigma_{yx}$ in the region where $\sigma_{xx}$ changes rapidly. We thus conclude that our mechanism provides a convincing explanation for the observed $\sigma_{yx}$ enhancement in our heterostructure samples, where amorphous and crystalline phases behave as if they are two different materials and are physically mixed.

Overall, the experimental results strongly support the theoretical model. The correlated changes of AHE and ANE and the microstructure demonstrate that the enhancement arises from the formation of island-like domains, which generate meandering current paths. A recent report also demonstrated that nanostructure engineering enhances transverse thermoelectric conversion~\cite{Gautam24NatCommun,PARK25ActaMater}; because such nanostructure engineering effectively creates composite domains similar to our system, our mechanism may provide a unified explanation. These consistent observations across different materials confirm the broad applicability of the proposed enhancement scheme in TT.

{\it Conclusions.---}The universality of the proposed approach arises from its independence from specific material parameters such as crystallographic orientation and electronic band structure. It can be applied to any binary physical mixtures that satisfy inequalities in the longitudinal and transverse conductivities between material A and B (i.e., $\gamma_{xx}^{(\text{A})} < \gamma_{xx}^{(\text{B})}$ and $\gamma_{yx}^{(\text{A})} > \gamma_{yx}^{(\text{B})}$), where hetero-domains generate meandering current paths. This principle enables creation of new materials with large TT simply by mixing existing components and may even surpass state-of-the-art systems. Although demonstrated for the anomalous Hall and Nernst effects, the concept is expected to apply to other forms of transverse flux.

{\it Note added---}After submission, our composite-based approach was also applied to the anomalous Ettingshausen effect, that is, transverse charge-to-heat conversion, and validated across over 150 samples with diverse compositions and annealing conditions~\cite{Park25arXiv}. This further supports the universality and tunability of the proposed mechanism and suggests its potential relevance to other types of flux, such as ionic or phononic Hall transport in heterogeneous conductors.

{\it Acknowledgments.---}The authors thank Proterial Korea, Ltd. for providing the commercial amorphous metals (2605SA1) for research purposes. The authors thank Ki Mun Bang for help with figures and Ji-Hyun Seong, Ho-Jun Gang, Jung-Hun Kim, and Eun-Ho Lee for their technical support. This work was supported by Samsung Research Funding \& Incubation Center of Samsung Electronics under Project Number SRFC-MA2002-02, National Research Foundation of Korea (NRF) grant funded by the Korea government (MSIT) RS-2024-00345022 and No. NRF-2022M3C1A3091988, Samsung Science and Technology Foundation grant BA-1501-51, National Research Foundation of Korea (NRF) grant funded by the Korean government (MSIT) RS-2024-00410027.

%\balance
\bibliography{paper_arXiv}

@misc{suppl_ref,
note = {See {S}upplemental {M}aterial for further details, which includes Refs.~\cite{Bang24ApplEnergy,Minor87JMS,McHenry03ScrMat,Zhou15PNSM,Tanimoto22JAC,Li23ActaMater,Rana03MCP,Watzman16PRB,Dobrosavljevic92PRL,Molinari23ACSAEM,Li21Matter,Lu15PRB,Lee85RMP,Uchida22Joule,Boona21JAP,Ikhlas17NatPhys,Sakai20Nature,Sakai18NatPhys,Pan22NatMater,He21Joule,Fu18EES,Guin19NPGAM,Guin19AdvMat,Mende21AdvSci,Zhou20AdvFuncMater,Zhou20APE,Xu20PRB,Ding19PRX,Zhou21NatMater,Secco17PEPI,Hsin17Nanotech,Buschinger97PhysicaB,Hamada21APL}.}
}

@article{Nagaosa10RMP,
  title = {Anomalous {H}all effect},
  author = {Nagaosa, Naoto and Sinova, Jairo and Onoda, Shigeki and MacDonald, A. H. and Ong, N. P.},
  journal = {Rev. Mod. Phys.},
  volume = {82},
  issue = {2},
  pages = {1539--1592},
  numpages = {0},
  year = {2010},
  month = {May},
  publisher = {American Physical Society},
  doi = {10.1103/RevModPhys.82.1539},
  url = {https://link.aps.org/doi/10.1103/RevModPhys.82.1539}
}

@article{Sinova15RMP,
  title = {Spin {H}all effects},
  author = {Sinova, Jairo and Valenzuela, Sergio O. and Wunderlich, J. and Back, C. H. and Jungwirth, T.},
  journal = {Rev. Mod. Phys.},
  volume = {87},
  issue = {4},
  pages = {1213--1260},
  numpages = {47},
  year = {2015},
  month = {Oct},
  publisher = {American Physical Society},
  doi = {10.1103/RevModPhys.87.1213},
  url = {https://link.aps.org/doi/10.1103/RevModPhys.87.1213}
}

@article{Chang23RMP,
  title = {Colloquium: {Q}uantum anomalous {H}all effect},
  author = {Chang, Cui-Zu and Liu, Chao-Xing and MacDonald, Allan H.},
  journal = {Rev. Mod. Phys.},
  volume = {95},
  issue = {1},
  pages = {011002},
  numpages = {33},
  year = {2023},
  month = {Jan},
  publisher = {American Physical Society},
  doi = {10.1103/RevModPhys.95.011002},
  url = {https://link.aps.org/doi/10.1103/RevModPhys.95.011002}
}

@article{Xiao10RMP,
  title = {Berry phase effects on electronic properties},
  author = {Xiao, Di and Chang, Ming-Che and Niu, Qian},
  journal = {Rev. Mod. Phys.},
  volume = {82},
  issue = {3},
  pages = {1959--2007},
  numpages = {0},
  year = {2010},
  month = {Jul},
  publisher = {American Physical Society},
  doi = {10.1103/RevModPhys.82.1959},
  url = {https://link.aps.org/doi/10.1103/RevModPhys.82.1959}
}

@article{Lee85RMP,
  title = {Disordered electronic systems},
  author = {Lee, Patrick A. and Ramakrishnan, T. V.},
  journal = {Rev. Mod. Phys.},
  volume = {57},
  issue = {2},
  pages = {287--337},
  numpages = {0},
  year = {1985},
  month = {Apr},
  publisher = {American Physical Society},
  doi = {10.1103/RevModPhys.57.287},
  url = {https://link.aps.org/doi/10.1103/RevModPhys.57.287}
}

@article{Kirkpatrick73RMP,
  title = {Percolation and Conduction},
  author = {Kirkpatrick, Scott},
  journal = {Rev. Mod. Phys.},
  volume = {45},
  issue = {4},
  pages = {574--588},
  numpages = {0},
  year = {1973},
  month = {Oct},
  publisher = {American Physical Society},
  doi = {10.1103/RevModPhys.45.574},
  url = {https://link.aps.org/doi/10.1103/RevModPhys.45.574}
}

@article{Karplus54PR,
  title = {{H}all Effect in Ferromagnetics},
  author = {Karplus, Robert and Luttinger, J. M.},
  journal = {Phys. Rev.},
  volume = {95},
  issue = {5},
  pages = {1154--1160},
  numpages = {0},
  year = {1954},
  month = {Sep},
  publisher = {American Physical Society},
  doi = {10.1103/PhysRev.95.1154},
  url = {https://link.aps.org/doi/10.1103/PhysRev.95.1154}
}

@article{Zhang10PRL,
  title = {Topological Nature of the Phonon {H}all Effect},
  author = {Zhang, Lifa and Ren, Jie and Wang, Jian-Sheng and Li, Baowen},
  journal = {Phys. Rev. Lett.},
  volume = {105},
  issue = {22},
  pages = {225901},
  numpages = {4},
  year = {2010},
  month = {Nov},
  publisher = {American Physical Society},
  doi = {10.1103/PhysRevLett.105.225901},
  url = {https://link.aps.org/doi/10.1103/PhysRevLett.105.225901}
}

@article{Kane05PRL,
  title = {${Z}_{2}$ Topological Order and the Quantum Spin {H}all Effect},
  author = {Kane, C. L. and Mele, E. J.},
  journal = {Phys. Rev. Lett.},
  volume = {95},
  issue = {14},
  pages = {146802},
  numpages = {4},
  year = {2005},
  month = {Sep},
  publisher = {American Physical Society},
  doi = {10.1103/PhysRevLett.95.146802},
  url = {https://link.aps.org/doi/10.1103/PhysRevLett.95.146802}
}

@article{Katsura10PRL,
  title = {Theory of the Thermal {H}all Effect in Quantum Magnets},
  author = {Katsura, Hosho and Nagaosa, Naoto and Lee, Patrick A.},
  journal = {Phys. Rev. Lett.},
  volume = {104},
  issue = {6},
  pages = {066403},
  numpages = {4},
  year = {2010},
  month = {Feb},
  publisher = {American Physical Society},
  doi = {10.1103/PhysRevLett.104.066403},
  url = {https://link.aps.org/doi/10.1103/PhysRevLett.104.066403}
}

@article{Jungwirth02PRL,
  title = {Anomalous {H}all Effect in Ferromagnetic Semiconductors},
  author = {Jungwirth, T. and Niu, Qian and MacDonald, A. H.},
  journal = {Phys. Rev. Lett.},
  volume = {88},
  issue = {20},
  pages = {207208},
  numpages = {4},
  year = {2002},
  month = {May},
  publisher = {American Physical Society},
  doi = {10.1103/PhysRevLett.88.207208},
  url = {https://link.aps.org/doi/10.1103/PhysRevLett.88.207208}
}

@article{Haldane04PRL,
  title = {Berry Curvature on the {F}ermi Surface: Anomalous {H}all Effect as a Topological {F}ermi-Liquid Property},
  author = {Haldane, F. D. M.},
  journal = {Phys. Rev. Lett.},
  volume = {93},
  issue = {20},
  pages = {206602},
  numpages = {4},
  year = {2004},
  month = {Nov},
  publisher = {American Physical Society},
  doi = {10.1103/PhysRevLett.93.206602},
  url = {https://link.aps.org/doi/10.1103/PhysRevLett.93.206602}
}

@article{Lee21PRL,
  title = {Non-{F}ermi Liquids in Conducting Two-Dimensional Networks},
  author = {Lee, Jongjun M. and Oshikawa, Masaki and Cho, Gil Young},
  journal = {Phys. Rev. Lett.},
  volume = {126},
  issue = {18},
  pages = {186601},
  numpages = {7},
  year = {2021},
  month = {May},
  publisher = {American Physical Society},
  doi = {10.1103/PhysRevLett.126.186601},
  url = {https://link.aps.org/doi/10.1103/PhysRevLett.126.186601}
}

@article{Dobrosavljevic92PRL,
  title = {Kondo effect in disordered systems},
  author = {Dobrosavljevi\ifmmode \acute{c}\else \'{c}\fi{}, V. and Kirkpatrick, T. R. and Kotliar, B. G.},
  journal = {Phys. Rev. Lett.},
  volume = {69},
  issue = {7},
  pages = {1113--1116},
  numpages = {0},
  year = {1992},
  month = {Aug},
  publisher = {American Physical Society},
  doi = {10.1103/PhysRevLett.69.1113},
  url = {https://link.aps.org/doi/10.1103/PhysRevLett.69.1113}
}

@article{Miyasato07PRL,
  title = {Crossover Behavior of the Anomalous {H}all Effect and Anomalous {N}ernst Effect in Itinerant Ferromagnets},
  author = {Miyasato, T. and Abe, N. and Fujii, T. and Asamitsu, A. and Onoda, S. and Onose, Y. and Nagaosa, N. and Tokura, Y.},
  journal = {Phys. Rev. Lett.},
  volume = {99},
  issue = {8},
  pages = {086602},
  numpages = {4},
  year = {2007},
  month = {Aug},
  publisher = {American Physical Society},
  doi = {10.1103/PhysRevLett.99.086602},
  url = {https://link.aps.org/doi/10.1103/PhysRevLett.99.086602}
}

@article{Xiao06PRL,
  title = {Berry-Phase Effect in Anomalous Thermoelectric Transport},
  author = {Xiao, Di and Yao, Yugui and Fang, Zhong and Niu, Qian},
  journal = {Phys. Rev. Lett.},
  volume = {97},
  issue = {2},
  pages = {026603},
  numpages = {4},
  year = {2006},
  month = {Jul},
  publisher = {American Physical Society},
  doi = {10.1103/PhysRevLett.97.026603},
  url = {https://link.aps.org/doi/10.1103/PhysRevLett.97.026603}
}

@article{Albert15PRL,
  title = {Topological Properties of Linear Circuit Lattices},
  author = {Albert, Victor V. and Glazman, Leonid I. and Jiang, Liang},
  journal = {Phys. Rev. Lett.},
  volume = {114},
  issue = {17},
  pages = {173902},
  numpages = {6},
  year = {2015},
  month = {Apr},
  publisher = {American Physical Society},
  doi = {10.1103/PhysRevLett.114.173902},
  url = {https://link.aps.org/doi/10.1103/PhysRevLett.114.173902}
}

@article{Cheianov07PRL,
  title = {Random Resistor Network Model of Minimal Conductivity in Graphene},
  author = {Cheianov, Vadim V. and Fal'ko, Vladimir I. and Altshuler, Boris L. and Aleiner, Igor L.},
  journal = {Phys. Rev. Lett.},
  volume = {99},
  issue = {17},
  pages = {176801},
  numpages = {4},
  year = {2007},
  month = {Oct},
  publisher = {American Physical Society},
  doi = {10.1103/PhysRevLett.99.176801},
  url = {https://link.aps.org/doi/10.1103/PhysRevLett.99.176801}
}

@article{Ding19PRX,
  title = {Intrinsic Anomalous {N}ernst Effect Amplified by Disorder in a Half-Metallic Semimetal},
  author = {Ding, Linchao and Koo, Jahyun and Xu, Liangcai and Li, Xiaokang and Lu, Xiufang and Zhao, Lingxiao and Wang, Qi and Yin, Qiangwei and Lei, Hechang and Yan, Binghai and Zhu, Zengwei and Behnia, Kamran},
  journal = {Phys. Rev. X},
  volume = {9},
  issue = {4},
  pages = {041061},
  numpages = {9},
  year = {2019},
  month = {Dec},
  publisher = {American Physical Society},
  doi = {10.1103/PhysRevX.9.041061},
  url = {https://link.aps.org/doi/10.1103/PhysRevX.9.041061}
}

@article{Fu07PRB,
  title = {Topological insulators with inversion symmetry},
  author = {Fu, Liang and Kane, C. L.},
  journal = {Phys. Rev. B},
  volume = {76},
  issue = {4},
  pages = {045302},
  numpages = {17},
  year = {2007},
  month = {Jul},
  publisher = {American Physical Society},
  doi = {10.1103/PhysRevB.76.045302},
  url = {https://link.aps.org/doi/10.1103/PhysRevB.76.045302}
}

@article{Berger70PRB,
  title = {Side-Jump Mechanism for the {H}all Effect of Ferromagnets},
  author = {Berger, L.},
  journal = {Phys. Rev. B},
  volume = {2},
  issue = {11},
  pages = {4559--4566},
  numpages = {0},
  year = {1970},
  month = {Dec},
  publisher = {American Physical Society},
  doi = {10.1103/PhysRevB.2.4559},
  url = {https://link.aps.org/doi/10.1103/PhysRevB.2.4559}
}

@article{Onoda08PRB,
  title = {Quantum transport theory of anomalous electric, thermoelectric, and thermal {H}all effects in ferromagnets},
  author = {Onoda, Shigeki and Sugimoto, Naoyuki and Nagaosa, Naoto},
  journal = {Phys. Rev. B},
  volume = {77},
  issue = {16},
  pages = {165103},
  numpages = {20},
  year = {2008},
  month = {Apr},
  publisher = {American Physical Society},
  doi = {10.1103/PhysRevB.77.165103},
  url = {https://link.aps.org/doi/10.1103/PhysRevB.77.165103}
}

@article{Medina13PRB,
  title = {Networks of quantum wire junctions: A system with quantized integer {H}all resistance without vanishing longitudinal resistivity},
  author = {Medina, Jaime and Green, Dmitry and Chamon, Claudio},
  journal = {Phys. Rev. B},
  volume = {87},
  issue = {4},
  pages = {045128},
  numpages = {7},
  year = {2013},
  month = {Jan},
  publisher = {American Physical Society},
  doi = {10.1103/PhysRevB.87.045128},
  url = {https://link.aps.org/doi/10.1103/PhysRevB.87.045128}
}

@article{Watzman16PRB,
  title = {Magnon-drag thermopower and {N}ernst coefficient in {F}e, {C}o, and {N}i},
  author = {Watzman, Sarah J. and Duine, Rembert A. and Tserkovnyak, Yaroslav and Boona, Stephen R. and Jin, Hyungyu and Prakash, Arati and Zheng, Yuanhua and Heremans, Joseph P.},
  journal = {Phys. Rev. B},
  volume = {94},
  issue = {14},
  pages = {144407},
  numpages = {9},
  year = {2016},
  month = {Oct},
  publisher = {American Physical Society},
  doi = {10.1103/PhysRevB.94.144407},
  url = {https://link.aps.org/doi/10.1103/PhysRevB.94.144407}
}

@article{Lu15PRB,
  title = {Weak antilocalization and localization in disordered and interacting {W}eyl semimetals},
  author = {Lu, Hai-Zhou and Shen, Shun-Qing},
  journal = {Phys. Rev. B},
  volume = {92},
  issue = {3},
  pages = {035203},
  numpages = {13},
  year = {2015},
  month = {Jul},
  publisher = {American Physical Society},
  doi = {10.1103/PhysRevB.92.035203},
  url = {https://link.aps.org/doi/10.1103/PhysRevB.92.035203}
}

@article{Xu20PRB,
  title = {Anomalous transverse response of $\mathrm{Co}_{2}\mathrm{MnGa}$ and universality of the room-temperature ${\ensuremath{\alpha}}_{ij}^{A}/{\ensuremath{\sigma}}_{ij}^{A}$ ratio across topological magnets},
  author = {Xu, Liangcai and Li, Xiaokang and Ding, Linchao and Chen, Taishi and Sakai, Akito and Fauqu\'e, Beno\^{\i}t and Nakatsuji, Satoru and Zhu, Zengwei and Behnia, Kamran},
  journal = {Phys. Rev. B},
  volume = {101},
  issue = {18},
  pages = {180404},
  numpages = {5},
  year = {2020},
  month = {May},
  publisher = {American Physical Society},
  doi = {10.1103/PhysRevB.101.180404},
  url = {https://link.aps.org/doi/10.1103/PhysRevB.101.180404}
}

@article{Zhou23PRAppl,
  title = {Direct Electrical Probing of Anomalous {N}ernst Conductivity},
  author = {Zhou, Weinan and Miura, Asuka and Sakuraba, Yuya and Uchida, Kenichi},
  journal = {Phys. Rev. Appl.},
  volume = {19},
  issue = {6},
  pages = {064079},
  numpages = {7},
  year = {2023},
  month = {Jun},
  publisher = {American Physical Society},
  doi = {10.1103/PhysRevApplied.19.064079},
  url = {https://link.aps.org/doi/10.1103/PhysRevApplied.19.064079}
}

@article{Sakai20Nature,
  title = {Iron-based binary ferromagnets for transverse thermoelectric conversion},
  author = {Sakai, Akito and Minami, Susumu and Koretsune, Takashi and Chen, Taishi and Higo, Tomoya and Wang, Yangming and Nomoto, Takuya and Hirayama, Motoaki and Miwa, Shinji and Nishio-Hamane, Daisuke and Ishii, Fumiyuki and Arita, Ryotaro and Nakatsuji, Satoru},
  journal = {Nature (London)},
  volume = {581},
  issue = {7806},
  pages = {53-57},
  year = {2020},
  month = {May},
  doi = {10.1038/s41586-020-2230-z},
  url = {https://doi.org/10.1038/s41586-020-2230-z}
}

@article{Cai23Nature,
  title = {Signatures of fractional quantum anomalous {H}all states in twisted $\text{MoTe}_{2}$},
  author = {Cai, Jiaqi and Anderson, Eric and Wang, Chong and Zhang, Xiaowei and Liu, Xiaoyu and Holtzmann, William and Zhang, Yinong and Fan, Fengren and Taniguchi, Takashi and Watanabe, Kenji and Ran, Ying and Cao, Ting and Fu, Liang and Xiao, Di and Yao, Wang and Xu, Xiaodong},
  journal = {Nature (London)},
  volume = {622},
  issue = {7981},
  pages = {63-68},
  year = {2023},
  month = {Oct},
  doi = {10.1038/s41586-023-06289-w},
  url = {https://doi.org/10.1038/s41586-023-06289-w}
}

@article{Sakai18NatPhys,
  title = {Giant anomalous {N}ernst effect and quantum-critical scaling in a ferromagnetic semimetal},
  author = {Sakai, Akito and Mizuta, Yo Pierre and Nugroho, Agustinus Agung and Sihombing, Rombang and Koretsune, Takashi and Suzuki, Michi-To and Takemori, Nayuta and Ishii, Rieko and Nishio-Hamane, Daisuke and Arita, Ryotaro and Goswami, Pallab and Nakatsuji, Satoru},
  journal = {Nat. Phys.},
  volume = {14},
  issue = {11},
  pages = {1119-1124},
  year = {2018},
  month = {Nov},
  doi = {10.1038/s41567-018-0225-6},
  url = {https://doi.org/10.1038/s41567-018-0225-6}
}

@article{Ikhlas17NatPhys,
  title = {Large anomalous {N}ernst effect at room temperature in a chiral antiferromagnet},
  author = {Ikhlas, Muhammad and Tomita, Takahiro and Koretsune, Takashi and Suzuki, Michi-To and Nishio-Hamane, Daisuke and Arita, Ryotaro and Otani, Yoshichika and Nakatsuji, Satoru},
  journal = {Nat. Phys.},
  volume = {13},
  issue = {11},
  pages = {1085-1090},
  year = {2017},
  month = {Nov},
  doi = {10.1038/nphys4181},
  url = {https://doi.org/10.1038/nphys4181}
}

@article{Takahagi25NatPhys,
  title = {Observation of the transverse {T}homson effect},
  author = {Takahagi, Atsushi and Hirai, Takamasa and Alasli, Abdulkareem and Park, Sang J. and Nagano, Hosei and Uchida, Kenichi},
  journal = {Nat. Phys.},
  volume = {21},
  issue = {8},
  pages = {1283-1289},
  year = {2025},
  month = {Oct},
  doi = {10.1038/s41567-025-02936-3},
  url = {https://doi.org/10.1038/s41567-025-02936-3}
}

@article{Klitzing20NatRevPhys,
  title = {40 years of the quantum {H}all effect},
  author = {von Klitzing, Klaus and Chakraborty, Tapash and Kim, Philip and Madhavan, Vidya and Dai, Xi and McIver, James and Tokura, Yoshinori and Savary, Lucile and Smirnova, Daria and Rey, Ana Maria and Felser, Claudia and Gooth, Johannes and Qi, Xiaoliang},
  journal = {Nat. Rev. Phys.},
  volume = {2},
  issue = {8},
  pages = {397-401},
  year = {2020},
  month = {Aug},
  doi = {10.1038/s42254-020-0209-1},
  url = {https://doi.org/10.1038/s42254-020-0209-1}
}

@article{Bauer12NatMater,
  title = {Spin caloritronics},
  author = {Bauer, Gerrit E. W. and Saitoh, Eiji and van Wees, Bart J.},
  journal = {Nat. Mater.},
  volume = {11},
  issue = {5},
  pages = {391-399},
  year = {2012},
  month = {May},
  doi = {10.1038/nmat3301},
  url = {https://doi.org/10.1038/nmat3301}
}

@article{Zhou21NatMater,
  title = {Seebeck-driven transverse thermoelectric generation},
  author = {Zhou, Weinan and Yamamoto, Kaoru and Miura, Asuka and Iguchi, Ryo and Miura, Yoshio and Uchida, Kenichi and Sakuraba, Yuya},
  journal = {Nat. Mater.},
  volume = {11},
  issue = {5},
  pages = {463-467},
  year = {2021},
  month = {Apr},
  doi = {10.1038/s41563-020-00884-2},
  url = {https://doi.org/10.1038/s41563-020-00884-2}
}

@article{Pan22NatMater,
  title = {Giant anomalous {N}ernst signal in the antiferromagnet $\text{YbMnBi}_{2}$},
  author = {Pan, Yu and Le, Congcong and He, Bin and Watzman, Sarah J. and Yao, Mengyu and Gooth, Johannes and Heremans, Joseph P. and Sun, Yan and Felser, Claudia},
  journal = {Nat. Mater.},
  volume = {21},
  issue = {2},
  pages = {203-209},
  year = {2022},
  month = {Feb},
  doi = {10.1038/s41563-021-01149-2},
  url = {https://doi.org/10.1038/s41563-021-01149-2}
}

@article{Dieny20NatElectron,
  title = {Opportunities and challenges for spintronics in the microelectronics industry},
  author = {Dieny, B. and Prejbeanu, I. L. and Garello, K. and Gambardella, P. and Freitas, P. and Lehndorff, R. and Raberg, W. and Ebels, U. and Demokritov, S. O. and Akerman, J. and Deac, A. and Pirro, P. and Adelmann, C. and Anane, A. and  Chumak, A. V. and Hirohata, A. and Mangin, S. and Valenzuela, Sergio O. and Onbaşlı, M. Cengiz and d’Aquino, M. and Prenat, G. and Finocchio, G. and Lopez-Diaz, L. and Chantrell, R. and Chubykalo-Fesenko, O. and Bortolotti, P.},
  journal = {Nat Electron.},
  volume = {3},
  issue = {8},
  pages = {446-459},
  year = {2020},
  month = {Aug},
  doi = {10.1038/s41928-020-0461-5},
  url = {https://doi.org/10.1038/s41928-020-0461-5}
}

@article{Gautam24NatCommun,
  title = {Creation of flexible spin-caloritronic material with giant transverse thermoelectric conversion by nanostructure engineering},
  author = {Gautam, Ravi and Hirai, Takamasa and Alasli, Abdulkareem and Nagano, Hosei and Ohkubo, Tadakatsu and Uchida, Kenichi and Sepehri-Amin, Hossein},
  journal = {Nat. Commun.},
  volume = {15},
  issue = {1},
  pages = {2184},
  year = {2024},
  month = {Mar},
  doi = {10.1038/s41467-024-46475-6},
  url = {https://doi.org/10.1038/s41467-024-46475-6}
}

@article{Ketter25NatCommun,
  title = {Using resistor network models to predict the transport properties of solid-state battery composites},
  author = {Ketter, Lukas and Greb, Niklas and Bernges, Tim and Zeier, Wolfgang G.},
  journal = {Nat. Commun.},
  volume = {16},
  issue = {1},
  pages = {1411},
  year = {2025},
  month = {Feb},
  doi = {10.1038/s41467-025-56514-5},
  url = {https://doi.org/10.1038/s41467-025-56514-5}
}

@article{Guin19NPGAM,
  title = {Anomalous {N}ernst effect beyond the magnetization scaling relation in the ferromagnetic {H}eusler compound $\text{Co2MnGa}$},
  author = {Guin, Satya N. and Manna, Kaustuv and Noky, Jonathan and Watzman, Sarah J. and Fu, Chenguang and Kumar, Nitesh and Schnelle, Walter and Shekhar, Chandra and Sun, Yan and Gooth, Johannes  and Felser, Claudia},
  journal = {NPG Asia Mater.},
  volume = {11},
  issue = {1},
  pages = {16},
  year = {2019},
  month = {Apr},
  doi = {10.1038/s41427-019-0116-z},
  url = {https://doi.org/10.1038/s41427-019-0116-z}
}

@article{Onose10Science,
  author = {Y. Onose  and T. Ideue  and H. Katsura  and Y. Shiomi  and N. Nagaosa  and Y. Tokura },
  title = {Observation of the Magnon {H}all Effect},
  journal = {Science},
  volume = {329},
  number = {5989},
  pages = {297-299},
  year = {2010},
  doi = {10.1126/science.1188260},
  URL = {https://www.science.org/doi/abs/10.1126/science.1188260}
}

@article{Chang13Science,
  author = {Cui-Zu Chang and Jinsong Zhang  and Xiao Feng and Jie Shen and Zuocheng Zhang and Minghua Guo and Kang Li and Yunbo Ou and Pang Wei and Li-Li Wang and Zhong-Qing Ji and Yang Feng and Shuaihua Ji and Xi Chen and Jinfeng Jia and Xi Dai and Zhong Fang and Shou-Cheng Zhang and Ke He and Yayu Wang and Li Lu and Xu-Cun Ma and Qi-Kun Xue},
  title = {Experimental Observation of the Quantum Anomalous {H}all Effect in a Magnetic Topological Insulator},
  journal = {Science},
  volume = {340},
  number = {6129},
  pages = {167-170},
  year = {2013},
  doi = {10.1126/science.1234414},
  URL = {https://www.science.org/doi/abs/10.1126/science.1234414}
}

@article{Konig07Science,
  author = {Markus König  and Steffen Wiedmann  and Christoph Brüne  and Andreas Roth  and Hartmut Buhmann  and Laurens W. Molenkamp  and Xiao-Liang Qi  and Shou-Cheng Zhang },
  title = {Quantum Spin {H}all Insulator State in $\text{HgTe}$ Quantum Wells},
  journal = {Science},
  volume = {318},
  number = {5851},
  pages = {766-770},
  year = {2007},
  doi = {10.1126/science.1148047},
  URL = {https://www.science.org/doi/abs/10.1126/science.1148047}
}

@article{Guin19AdvMat,
  author = {Guin, Satya N. and Vir, Praveen and Zhang, Yang and Kumar, Nitesh and Watzman, Sarah J. and Fu, Chenguang and Liu, Enke and Manna, Kaustuv and Schnelle, Walter and Gooth, Johannes and Shekhar, Chandra and Sun, Yan and Felser, Claudia},
  title = {Zero-Field {N}ernst Effect in a Ferromagnetic {K}agome-Lattice {W}eyl-Semimetal $\text{Co}_{3}\text{Sn}_{2}\text{S}_{2}$},
  journal = {Adv. Mater.},
  volume = {31},
  number = {25},
  pages = {1806622},
  doi = {https://doi.org/10.1002/adma.201806622},
  url = {https://advanced.onlinelibrary.wiley.com/doi/abs/10.1002/adma.201806622},
  year = {2019}
}

@article{Zhou20AdvFuncMater,
  author = {Zhou, Wu-Xing and Cheng, Yuan and Chen, Ke-Qiu and Xie, Guofeng and Wang, Tian and Zhang, Gang},
  title = {Thermal Conductivity of Amorphous Materials},
  journal = {Adv. Funct. Mater.},
  volume = {30},
  number = {8},
  pages = {1903829},
  doi = {10.1002/adfm.201903829},
  url = {https://doi.org/10.1002/adfm.201903829},
  year = {2020}
}

@article{Mende21AdvSci,
  author = {Mende, Felix and Noky, Jonathan and Guin, Satya N. and Fecher, Gerhard H. and Manna, Kaustuv and Adler, Peter and Schnelle, Walter and Sun, Yan and Fu, Chenguang and Felser, Claudia},
  title = {Large Anomalous {H}all and {N}ernst Effects in High {C}urie-Temperature Iron-Based {H}eusler Compounds},
  journal = {Adv. Sci.},
  volume = {8},
  number = {17},
  pages = {2100782},
  doi = {10.1002/advs.202100782},
  url = {https://doi.org/10.1002/advs.202100782},
  year = {2021}
}

@article{Rosch21ATS,
  author = {Rösch, Andres Georg and Giunta, Fabian and Mallick, Md. Mofasser and Franke, Leonard and Gall, André and Aghassi-Hagmann, Jasmin and Schmalian, Jörg and Lemmer, Uli},
  title = {Improved Electrical, Thermal, and Thermoelectric Properties Through Sample-to-Sample Fluctuations in Near-Percolation Threshold Composite Materials},
  journal = {Adv. Theor. Simul.},
  volume = {4},
  number = {6},
  pages = {2000284},
  doi = {10.1002/adts.202000284},
  url = {https://doi.org/10.1002/adts.202000284},
  year = {2021}
}

@article{Li23ActaMater,
  author = {Le Li and Zhenghao Chen and Shogo Kuroiwa and Mitsuhiro Ito and Koretaka Yuge and Kyosuke Kishida and Hisanori Tanimoto and Yue Yu and Haruyuki Inui and Easo P. George},
  title = {Evolution of short-range order and its effects on the plastic deformation behavior of single crystals of the equiatomic $\text{Cr}-\text{Co}-\text{Ni}$ medium-entropy alloy},
  journal = {Acta Mater.},
  volume = {243},
  pages = {118537},
  year = {2023},
  doi = {10.1016/j.actamat.2022.118537},
  url = {https://doi.org/10.1016/j.actamat.2022.118537}
}

@article{PARK25ActaMater,
  author = {Sang J. Park and Rajkumar Modak and Ravi Gautam and Abdulkareem Alasli and Takamasa Hirai and Fuyuki Ando and Hosei Nagano and Hossein Sepehri-Amin and Kenichi Uchida},
  title = {Designing flexible magnetic materials for zero-magnetic-field operation of the anomalous {N}ernst effect},
  journal = {Acta Mater.},
  volume = {301},
  pages = {121422},
  year = {2025},
  doi = {10.1016/j.actamat.2025.121422},
  url = {https://doi.org/10.1016/j.actamat.2025.121422}
}

@article{Li21Matter,
  author = {Li, Xiang and Li, Peng and D.-H. Hou, Vincent and DC, Mahendra and Nien, Chih-Hung and Xue, Fen and Yi, Di and Bi, Chong and Lee, Chien-Min and Lin, Shy-Jay and Tsai, Wilman and Suzuki, Yuri and X. Wang, Shan},
  title = {Large and robust charge-to-spin conversion in sputtered conductive $\text{WTe}_{x}$ with disorder},
  journal = {Matter},
  volume = {4},
  pages = {1639-1653},
  year = {2021},
  doi = {10.1016/j.matt.2021.02.016},
  url = {https://doi.org/10.1016/j.matt.2021.02.016}
}

@article{Park24Matter,
  author = {Sang J. Park and Phuoc {Cao Van} and Min-Gu Kang and Hyeon-Jung Jung and Gi-Yeop Kim and Si-Young Choi and Jung-Woo Yoo and Byong-Guk Park and Se Kwon Kim and Jong-Ryul Jeong and Hyungyu Jin},
  title = {Enhancing spin pumping by nonlocal manipulation of magnon temperature},
  journal = {Matter},
  volume = {7},
  number = {12},
  pages = {4332-4341},
  year = {2024},
  doi = {10.1016/j.matt.2024.08.023},
  url = {https://doi.org/10.1016/j.matt.2024.08.023}
}

@article{Chalker88JPC,
  author = {J T Chalker and P D Coddington},
  title = {Percolation, quantum tunnelling and the integer {H}all effect},
  journal = {J. Phys. C},
  year = {1988},
  month = {may},
  volume = {21},
  number = {14},
  pages = {2665},
  doi = {10.1088/0022-3719/21/14/008},
  url = {https://doi.org/10.1088/0022-3719/21/14/008}
}

@article{Boona21JAP,
  author = {Boona, Stephen R. and Jin, Hyungyu and Watzman, Sarah},
  title = {Transverse thermal energy conversion using spin and topological structures},
  journal = {J. Appl. Phys.},
  volume = {130},
  number = {17},
  pages = {171101},
  year = {2021},
  month = {11},
  doi = {10.1063/5.0062559},
  url = {https://doi.org/10.1063/5.0062559}
}

@article{Bergman91JAP,
  author = {Bergman, David J. and Levy, Ohad},
  title = {Thermoelectric properties of a composite medium},
  journal = {J. Appl. Phys.},
  volume = {70},
  number = {11},
  pages = {6821-6833},
  year = {1991},
  month = {12},
  doi = {10.1063/1.349830},
  url = {https://doi.org/10.1063/1.349830}
}

@article{Minor87JMS,
  author = {Minor, W. and Schönfeld, B. and Lebech, B. and Buras, B. and Dmowski, W.},
  title = {Crystallization of {F}e-{S}i-{B} metallic glasses studied by {X}-ray synchrotron radiation},
  journal = {J. Mater. Sci.},
  volume = {22},
  pages = {4144-4152},
  year = {1987},
  month = {Nov},
  doi = {10.1007/BF01133371},
  url = {https://doi.org/10.1007/BF01133371}
}

@article{Molinari23ACSAEM,
  author = {Molinari, Alan and Balduini, Federico and Rocchino, Lorenzo and Wawrzyńczak, Rafał and Sousa, Marilyne and Bui, Holt and Lavoie, Christian and Stanic, Vesna and Jordan-Sweet, Jean and Hopstaken, Marinus and Tchoumakov, Serguei and Franca, Selma and Gooth, Johannes and Fratini, Simone and Grushin, Adolfo G. and Zota, Cezar and Gotsmann, Bernd and Schmid, Heinz},
  title = {Disorder-Induced Magnetotransport Anomalies in Amorphous and Textured $\text{Co}_{1–x}\text{Si}_{x}$ Semimetal Thin Films},
  journal = {J. Mater. Sci.},
  volume = {5},
  pages = {2624-2637},
  year = {2023},
  month = {May},
  doi = {10.1021/acsaelm.3c00095},
  url = {https://doi.org/10.1021/acsaelm.3c00095}
}

@article{Nakatani24APL,
    author = {Nakatani, Tomoya and Kulkarni, Prabhanjan D. and Suto, Hirofumi and Masuda, Keisuke and Iwasaki, Hitoshi and Sakuraba, Yuya},
    title = {Perspective on nanoscale magnetic sensors using giant anomalous {H}all effect in topological magnetic materials for read head application in magnetic recording},
    journal = {Appl. Phys. Lett.},
    volume = {124},
    number = {7},
    pages = {070501},
    year = {2024},
    month = {02},
    issn = {0003-6951},
    doi = {10.1063/5.0191974},
    url = {https://doi.org/10.1063/5.0191974}
}

@article{Bang24ApplEnergy,
  title = {Large transverse thermopower in shape-engineered tilted leg thermopile},
  journal = {Appl. Energy},
  volume = {368},
  pages = {123222},
  year = {2024},
  doi = {10.1016/j.apenergy.2024.123222},
  url = {https://doi.org/10.1016/j.apenergy.2024.123222},
  author = {Ki Mun Bang and Sang J. Park and Hyun Yu and Hyungyu Jin}
}

@article{Park25ATE,
title = {High heat-flux sensitivity of the planar coil device based on the anomalous {N}ernst effect},
journal = {Appl. Therm. Eng.},
volume = {265},
pages = {125555},
year = {2025},
doi = {10.1016/j.applthermaleng.2025.125555},
url = {https://doi.org/10.1016/j.applthermaleng.2025.125555},
author = {Sang J. Park and Ki Mun Bang and Jinho Park and Hyungyu Jin}
}

@article{Boona14EES,
  author = {Boona, Stephen R. and Myers, Roberto C. and Heremans, Joseph P.},
  title  = {Spin caloritronics},
  journal  = {Energy Environ. Sci.},
  year  = {2014},
  volume  = {7},
  issue  = {3},
  pages  = {885-910},
  publisher  = {The Royal Society of Chemistry},
  doi  = {10.1039/C3EE43299H},
  url  = {http://dx.doi.org/10.1039/C3EE43299H}
}

@Article{Fu18EES,
  author={Fu, Chenguang and Guin, Satya N and Watzman, Sarah J and Li, Guowei and Liu, Enke and Kumar, Nitesh and S{\"u}$\beta$, Vicky and Schnelle, Walter and Auffermann, Gudrun and Shekhar, Chandra and others},
  title  = {Large {N}ernst power factor over a broad temperature range in polycrystalline {W}eyl semimetal {N}b{P}},
  journal  = {Energy Environ. Sci.},
  year  = {2018},
  volume  = {11},
  issue  = {10},
  pages  = {2813-2820},
  publisher  = {The Royal Society of Chemistry},
  doi  = {10.1039/C8EE02077A},
  url  = {http://dx.doi.org/10.1039/C8EE02077A}
}

@article{Uchida22Joule,
  title = {Thermoelectrics: From longitudinal to transverse},
  journal = {Joule},
  volume = {6},
  number = {10},
  pages = {2240-2245},
  year = {2022},
  doi = {10.1016/j.joule.2022.08.016},
  url = {https://doi.org/10.1016/j.joule.2022.08.016},
  author = {Kenichi Uchida and Joseph P. Heremans}
}

@article{He21Joule,
  title = {Large magnon-induced anomalous {N}ernst conductivity in single-crystal MnBi},
  journal = {Joule},
  volume = {5},
  number = {11},
  pages = {3057},
  year = {2021},
  doi = {10.1016/j.joule.2021.08.007},
  url = {https://doi.org/10.1016/j.joule.2021.08.007},
  author = {He, Bin and {\c{S}}ahin, C{\"u}neyt and Boona, Stephen R and Sales, Brian C and Pan, Yu and Felser, Claudia and Flatt{\'e}, Michael E and Heremans, Joseph P}
}

@article{Smit58Physica,
  title = {The spontaneous {H}all effect in ferromagnetics {II}},
  journal = {Physica},
  volume = {24},
  number = {1},
  pages = {39-51},
  year = {1958},
  doi = {10.1016/S0031-8914(58)93541-9},
  url = {https://doi.org/10.1016/S0031-8914(58)93541-9},
  author = {J. Smit}
}

@article{Buschinger97PhysicaB,
  title = {Transport properties of {F}e{S}i},
  journal = {Physica B (Amsterdam)},
  volume = {230-232},
  pages = {784-786},
  year = {1997},
  doi = {10.1016/S0921-4526(96)00839-3},
  url = {https://doi.org/10.1016/S0921-4526(96)00839-3},
  author = {B. Buschinger and C. Geibel and F. Steglich and D. Mandrus and D. Young and J.L. Sarrao and Z. Fisk}
}

@article{Hebel64PhysLett,
  title = {Interband transitions and band structure of a {B}i{S}b alloy},
  journal = {Phys. Lett.},
  volume = {10},
  number = {3},
  pages = {273-275},
  year = {1964},
  doi = {10.1016/0031-9163(64)90498-6},
  url = {https://doi.org/10.1016/0031-9163(64)90498-6},
  author = {L.C. Hebel and G.E. Smith}
}

@article{Stauffer79PhysRep,
  title = {Scaling theory of percolation clusters},
  journal = {Phys. Rep.},
  volume = {54},
  number = {1},
  pages = {1-74},
  year = {1979},
  doi = {10.1016/0370-1573(79)90060-7},
  url = {https://doi.org/10.1016/0370-1573(79)90060-7},
  author = {D. Stauffer}
}

@article{Monetti93PBC,
  title = {Percolation on the square lattice in $a\text{L}×\text{M}$ geometry},
  journal = {Z. Phys. B},
  volume = {90},
  pages = {351-355},
  year = {1993},
  doi = {10.1007/BF01433059},
  url = {https://doi.org/10.1007/BF01433059},
  author = {Monetti, R. A. and Albano, E. V.}
}

@article{Rana03MCP,
  title = {Electrical resistivity behavior in {N}i–25 at. {C}r alloy},
  journal = {Mater. Chem. Phys.},
  volume = {80},
  number = {1},
  pages = {228-231},
  year = {2003},
  doi = {10.1016/S0254-0584(02)00468-6},
  url = {https://doi.org/10.1016/S0254-0584(02)00468-6},
  author = {Anwar Manzoor Rana and Abdul Faheem Khan and Amer Abbas and M.Iqbal Ansari}
}

@article{Hirohata20JMMM,
  title = {Review on spintronics: Principles and device applications},
  journal = {J. Magn. Magn. Mater.},
  volume = {509},
  pages = {166711},
  year = {2020},
  doi = {10.1016/j.jmmm.2020.166711},
  url = {https://doi.org/10.1016/j.jmmm.2020.166711},
  author = {Atsufumi Hirohata and Keisuke Yamada and Yoshinobu Nakatani and Ioan-Lucian Prejbeanu and Bernard Diény and Philipp Pirro and Burkard Hillebrands}
}

@article{Tanimoto22JAC,
  title = {Electrical resistivity and short-range order in rapid-quenched {C}r{M}n{F}e{C}o{N}i high-entropy alloy},
  journal = {J. Alloys Compd.},
  volume = {896},
  pages = {163059},
  year = {2022},
  doi = {10.1016/j.jallcom.2021.163059},
  url = {https://doi.org/10.1016/j.jallcom.2021.163059},
  author = {Hisanori Tanimoto and Ryo Hozumi and Mari Kawamura}
}

@article{McHenry03ScrMat,
  title = {The kinetics of nanocrystallization and microstructural observations in {FINEMET}, {NANOPERM} and {HITPERM} nanocomposite magnetic materials},
  journal = {Scr. Mater.},
  volume = {48},
  number = {7},
  pages = {881-887},
  year = {2003},
  doi = {10.1016/S1359-6462(02)00597-3},
  url = {https://doi.org/10.1016/S1359-6462(02)00597-3},
  author = {M.E. McHenry and F. Johnson and H. Okumura and T. Ohkubo and V.R.V. Ramanan and D.E. Laughlin}
}

@article{Zhou15PNSM,
  title = {Influence of clusters in melt on the subsequent glass-formation and crystallization of {F}e–{S}i–{B} metallic glasses},
  journal = {Prog. Nat. Sci.},
  volume = {25},
  number = {2},
  pages = {137-140},
  year = {2015},
  doi = {10.1016/j.pnsc.2015.02.002},
  url = {https://doi.org/10.1016/j.pnsc.2015.02.002},
  author = {Shaoxiong Zhou and Bangshao Dong and Rui Xiang and Guangqiang Zhang and Jingyu Qin and Xiufang Bian}
}

@article{Hsin17Nanotech,
  doi = {10.1088/1361-6528/aa904a},
  url = {https://doi.org/10.1088/1361-6528/aa904a},
  year = {2017},
  month = {nov},
  volume = {28},
  number = {48},
  pages = {485702},
  author = {Hsin, Cheng-Lun and Liu, Yu-Ting and Tsai, Yue-Yun},
  title = {Suppressed {U}mklapp scattering of $\beta-\text{FeSi}_{2}$ thin film and single crystalline nanowires},
  journal = {Nanotechnology}
}

@article{Zhou20APE,
  doi = {10.35848/1882-0786/ab79fe},
  url = {https://doi.org/10.35848/1882-0786/ab79fe},
  year = {2020},
  month = {mar},
  volume = {13},
  number = {4},
  pages = {043001},
  author = {Zhou, Weinan and Sakuraba, Yuya},
  title = {Heat flux sensing by anomalous {N}ernst effect in {F}e-{A}l thin films on a flexible substrate},
  journal = {Appl. Phys. Express}
}

@article{Hamada21APL,
  title = {Anomalous {N}ernst effect in {F}e-{S}i alloy films},
  author = {Yuki Hamada and Yuichiro Kurokawa and Tomoki Yamauchi and Hiroki Hanamoto and Hiromi Yuasa},
  journal = {Appl. Phys. Lett.},
  volume = {119},
  number = {15},
  pages = {152404},
  year = {2021},
  month = {Oct},
  doi = {10.1063/5.0062637},
  url = {https://doi.org/10.1063/5.0062637}
}

@article{Secco17PEPI,
  title = {Thermal conductivity and {S}eebeck coefficient of {F}e and {F}e-{S}i alloys: Implications for variable {L}orenz number},
  journal = {Phys. Earth Planet. Inter.},
  volume = {265},
  pages = {23-34},
  year = {2017},
  doi = {10.1016/j.pepi.2017.01.005},
  url = {https://doi.org/10.1016/j.pepi.2017.01.005},
  author = {Richard A. Secco}
}

@article{Tumanski13PE,
  author = {Tumanski, S.},
  year = {2013},
  month = {01},
  pages = {1-12},
  title = {Modern magnetic field sensors - a review},
  volume = {89},
  journal = {Prz. Elektrotech.}
}

@misc{Park25arXiv,
  title={High-throughput development of flexible amorphous materials showing large anomalous {N}ernst effect via automatic annealing and thermoelectric imaging, (to be published)}, 
  author={Park, S. J. and Gautam, R. and Alasli, A. and Hirai, T. and Ando, F. and Nagano, H. and Sepehri-Amin, H. and Uchida, K.}
}

% ---------------------------
% End Matter section
% ---------------------------
\onecolumngrid
\begin{center}
  \textbf{End Matter}
\end{center}

\twocolumngrid
\clearpage

{\it Details on the two-dimensional network model.---}Our two-dimensional square network model is similar to those introduced to simulate longitudinal thermoelectric transport~\cite{Rosch21ATS}, but includes additional degrees of freedom to capture TT. The network consists of sites with intrinsic transport properties connected by wires [Figs.~\ref{fig1}(a) and \ref{fig1}(b)]. We assume the linear response theory, in which the steady current $I_{i}^{\alpha}$ flowing out of a site $\alpha$ through the wire $i$ is related to the voltage $V_{j}^{\alpha}$ by the classical Ohm's law $I_{i}^{\alpha} = \sum_{j}G_{ij}^{\alpha}V_{j}^{\alpha}$. Charge conservation at each site imposes Kirchhoff's junction law, $\sum_{i}G_{ij}^{\alpha} = 0$. Because the currents should be invariant even if the voltages are all shifted by a constant, $\sum_{j}G_{ij}^{\alpha}=0$ also holds. When a magnetic field or magnetization is applied along the $\hat{z}$ (out-of-plane) direction, the system retains $C_{4z}$ rotational symmetry, while time-reversal and in-plane mirror symmetries are broken. With the $C_{4z}$ symmetry imposed, the two sum rules constrain the 16 components of the local conductance tensor $G_{ij}^{\alpha}$ to combinations of three independent parameters: $G_{\text{S}}^{\alpha}$, $G_{\text{S}}^{\prime \alpha}$, and $G_{\text{A}}^{\alpha}$. Here, $G_{\text{A}}^{\alpha}$ arises from the breaking of time-reversal symmetry and characterizes the antisymmetric part of $G_{ij}^{\alpha}$. The symmetric part of $G_{ij}^{\alpha}$ requires two independent components, $G_{\text{S}}^{\alpha}$ and $G_{\text{S}}^{\prime \alpha}$, chosen to characterize $G_{ii}^{\alpha}$ and $G_{i,i+2}^{\alpha}$, respectively. Since it is natural to assume that $G_{\text{S}}^{\alpha} \neq 0$, we introduce $c^{\alpha} = -G_{\text{S}}^{\prime \alpha}/G_{\text{S}}^{\alpha}$. The local conductance tensor $G_{ij}^{\alpha}$ is then expressed by Eq.~\eqref{eq:1}. Finally, we construct a finite $N_{x} \times N_{y}$ square network with four edges by connecting the sites.

{\it Numerical calculation.---}The transport of the entire network system is determined by the $G_{ij}^{\alpha}$ of each site and the boundary condition. It is described by the linear equation $\textbf{I} = \textbf{G}\textbf{V}$ with constraints, where $\textbf{G} = \bigoplus^{N}_{\alpha = 1}G^{\alpha}$, $\textbf{I} = (I^{1}, I^{2}, ..., I^{N})^{\textrm{T}}$, and $\textbf{V} = (V^{1}, V^{2}, ..., V^{N})^{\textrm{T}}$. Here, $\bigoplus$ denotes the direct sum of the $4 \times 4$ matrices $G^{\alpha}$, and $N = N_{x}N_{y}$. The boundary condition is set on the four edges. The currents passing through the left and right edges of the system are fixed to be zero, while the currents injected to the bottom edge and the currents extracted from the top edge are set to be uniform. To solve this linear equation with the boundary condition, we adopt a method similar to that used in previous studies on the honeycomb network model~\cite{Medina13PRB,Lee21PRL}. For pairs of nearest-neighbor sites $\alpha$ and $\beta$, we introduce the constraints $V_{i}^{\alpha} = V_{j}^{\beta}$ and $I_{i}^{\alpha} = -I_{j}^{\beta}$, where wire $i$ of site $\alpha$ and wire $j$ of site $\beta$ are connected to each other. Finally, we utilize the freedom of simultaneous voltage shifts of all sites and set the voltages at a chosen site zero. Then, we perform a numerical calculation to solve this reduced linear equation and obtain the voltages on the edges. In order to simulate the values measured in actual experiments, we repeat the calculation for the opposite direction of the magnetization or magnetic field and corrected the voltage. $\gamma_{xx}$ and $\gamma_{yx}$ of the entire system are calculated from these voltages and input currents. In the absence of a physical mixture, $\gamma_{xx}$ and $\gamma_{xy}$ in a sufficiently large network is independent of the network size and is entirely determined by the conductivity parameters, $G_{\text{S}}^{\alpha}$, $c^{\alpha}$, and $G_{\text{A}}^{\alpha}$ (Note S1 in Supplemental Material~\cite{suppl_ref}).

The results of Fig.~\ref{fig1}(d) in the main text was obtained for $N_{x} = N_{y} = 40$. To obtain $\bar{\gamma}_{yx}$ of the random network in Fig.~\ref{fig1}(d) (red points), we averaged $\gamma_{yx}$ for 100 randomly generated domain configurations for a given $P^{(B)}$. The results of Fig.~\ref{fig1}(e) is obtained for a square network with $N_{x} = N_{y} = 20$. To determine whether the mixture-induced enhancement occurs at each point of the diagram, we selected 15 values of portion $P^{(B)}$ at equal intervals and calculated $\bar{\gamma}_{yx}$ for each $P^{(B)}$. To focus on the effect of the site parameters, we repeated the calculations for the same $P^{(B)}$ and found $0.2-\sigma$ interval from its average value. We assess whether the enhancement occurs for a given set of $\gamma_{yx}^{({\rm A})}$ and $\gamma_{yx}^{({\rm B})}$ by checking if there exists a range of $P^{(B)}$ at which the infimum of its interval is larger than the values of $\gamma_{yx}$ at $P^{(B)}=0$ and $1$.

%%%%%%%%%%%%%%%%<Figure>%%%%%%%%%%%%%%%%
\begin{figure}[t!]
\includegraphics[width=245pt]{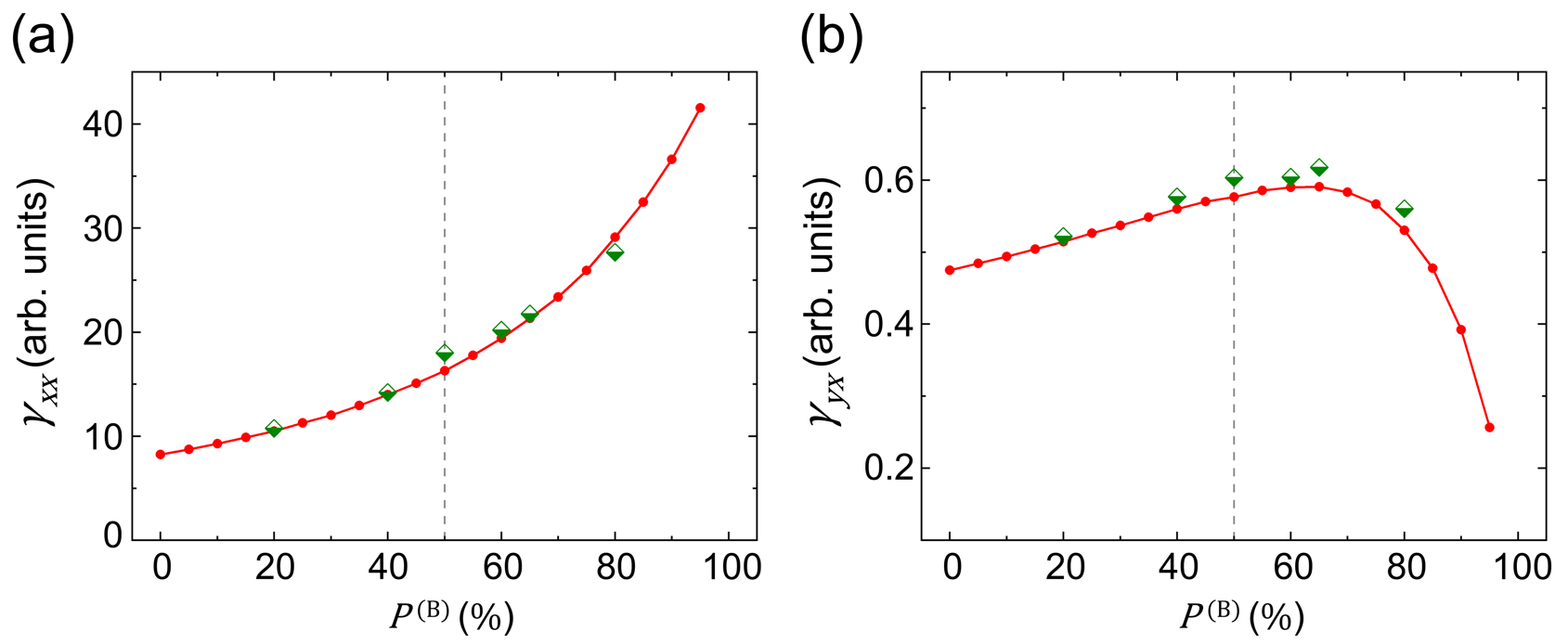}
\caption{\label{fig5}(Color online) Calculated (a) $\gamma_{xx}$ and (b) $\gamma_{yx}$ for moderate contrast in transport properties (see End Matter). Red symbols denote (a) $\bar{\gamma}_{xx}$ and (b) $\bar{\gamma}_{yx}$, whereas green diamonds represent (a) $\gamma_{xx}$ and (b) $\gamma_{yx}$ for the island structure at $P^{(\text{B})}=50\%$ and modified structures from the island structure for $P^{(\text{B})} \neq 50\%$.}
\end{figure}
%%%%%%%%%%%%%%%%%%%%%%%%%%%%%%%%%%%%%%%%

{\it Effect of domain geometry on current flow.---}To demonstrate that the geometric structure of physical domains generates the meandering current paths, we examine the current and potential distributions of the island structures (inset in Figs.~\ref{fig1}(d) and~\ref{fig2}). We first consider the island structure where material A (amorphous, yellow) islands are embedded in material B (crystalline, cyan) [Fig.~\ref{fig2}(a)]. Without a physical mixture (blue dots in Fig.~\ref{fig1}(d)), the macroscopic potential varies linearly, and its gradient is determined solely by the ratio of longitudinal to transverse conductivity. Introducing island domains distorts this potential and causes the meandering current paths [Fig.~\ref{fig2}(b)], as the current tends to flow along the material B region, avoiding the material A islands (red arrows in Figs.\ref{fig2}(b) and \ref{fig2}(c)). To show this more clearly, we obtain the current flowing in site A1 (green circle in Fig.~\ref{fig2}(a)) and B1 (red circle in Fig.~\ref{fig2}(a)) from the potential distribution. The longitudinal current strength ($\vert I_{3} - I_{1} \vert$) at B1 is 25.04, 12.13 times larger than the longitudinal current at A1, while the ratio of the transverse current strength to the longitudinal current strength ($\vert I_{4} - I_{2} \vert / \vert I_{3} - I_{1} \vert$) is 0.60 at B1, 5.04 times smaller than at A1. To highlight the transverse flow of the current, we calculate $[V^\alpha_i-V^\alpha_i(\mathcal{T})]/2$ [Figs.~\ref{fig2}(c) and~\ref{fig2}(f)]. Here, $V^\alpha_i$ is the potential at wire $i$ of site $\alpha$ and $V^\alpha_i(\mathcal{T})$ is the corresponding quantity for a time-reversed system, where the signs of $G_{\rm A}^{({\rm A})}$ and $G_{\rm A}^{({\rm B})}$ are reversed. The island locally improves the asymmetry of the transverse potential, indicating an imbalance in the deflected current. These results are consistent with our analysis in the main text. A similar tendency is found when material B islands are embedded in material A [Fig.~\ref{fig2}(d)]. In this case, the current tends to minimize the portion in material A [Figs.~\ref{fig2}(e) and ~\ref{fig2}(f)], which still leads to meandering current paths.

{\it Note on parameter choice and experimental realism.---}The local conductance tensor parameters are chosen to reproduce the transport properties of each material. In the absence of a physical mixture, both $\gamma_{xx}$ and $\gamma_{yx}$ of a sufficiently large network are fully determined by these parameters (Note S1 in Supplemental Material~\cite{suppl_ref}). $\gamma_{xx}$ is determined by the combination of $G_{\text{S}}^{\alpha}$ and $c^{\alpha}$, while $\gamma_{yx}$ is proportional to $G_{\text{A}}^{\alpha}$. Thus, we first assign $\gamma_{xx}$ and $\gamma_{yx}$ of the materials A and B, then we select the corresponding conductance parameters accordingly. To clearly demonstrate that the physical mixture induces meandering current paths and thereby enhances TT, we set a large contrast between $\gamma_{xx}$ of materials A and B. For the results of Fig~\ref{fig1}(d), we choose ($G_{\text{S}}^{(\text{A})}$, $c^{(\text{A})}$, $G_{\text{A}}^{(\text{A})}$) = (5, 0.8, 0.6) and ($G_{\text{S}}^{(\text{B})}$, $c^{(\text{B})}$, $G_{\text{A}}^{(\text{B})}$) = (100, 0.9, 0.01). The result of Fig.~\ref{fig1}(e) is obtained by using the same conductance parameters, except for $G_{\text{A}}^{(\text{A})}$ and $G_{\text{A}}^{(\text{B})}$. Under this choice of parameters, the contrast in $\gamma_{yx}^{\alpha}$ ($\gamma_{yx}^{(\text{A})} > \gamma_{yx}^{(\text{B})}$) generally induces enhanced TT.

In many practical situations, the contrast in transport properties between two materials (or phases) is smaller than that used in our calculation [Fig.~\ref{fig1}(d)]. In our experiment, for instance, the ratios of the two phases' transport properties are $\gamma_{xx}^{(\text{B})}/\gamma_{xx}^{(\text{A})} = 1.9$ and $\gamma_{yx}^{(\text{A})}/\gamma_{yx}^{(\text{B})} = 3.1$, yielding a TT enhancement of 70\% (at $P^{(\text{B})} = 80\%$). We find that our model can reproduce the enhancement even for contrasts much more moderate than those used in Figs.~\ref{fig1} and \ref{fig2}. For example, Fig.~\ref{fig5} shows a 30\% enhancement (at $P^{(\text{B})} = 65\%$) obtained with conductance parameters satisfying $\gamma_{xx}^{(\text{B})}/\gamma_{xx}^{(\text{A})} = 5.0$ and $\gamma_{yx}^{(\text{B})}/\gamma_{yx}^{(\text{B})} = 1.7$. Here, we evaluate the transport properties of the crystalline phase at $P^{(\text{B})} = 95\%$, considering that in actual experiments the crystalline phase formed at high-temperature is not a perfectly homogeneous single crystal but consists of densely packed crystalline domains, as shown in Fig.~\ref{fig3}(d). The corresponding conductance parameters are ($G_{\text{S}}^{(\text{A})}$, $c^{(\text{A})}$, $G_{\text{A}}^{(\text{A})}$) = (15, 0.1, 0.95) and ($G_{\text{S}}^{(\text{B})}$, $c^{(\text{B})}$, $G_{\text{A}}^{(\text{B})}$) = (50, 0.9, 0.1). Our simplified model accounts only for the bending of current paths that arises when domains with distinct transport properties are spatially arranged while retaining their intrinsic characteristics. In real situations, carrier scattering at domain boundaries further amplifies the meandering current paths, implying that even stronger enhancements can occur under milder contrasts than our model presents. Such scattering effects could be incorporated by adding inter-site scattering terms in our model, which would improve the predictive capability for finding material combinations. This extension lies beyond the present scope, and we leave this for future work.

% ---------------------------
% Supplemental Material section
% ---------------------------

\onecolumngrid

\newpage

\appendix

\begin{center}
  \large{\textbf{Supplemental Material for ``Enhanced transverse electron transport via disordered composite formation"}}
\end{center}

\tableofcontents
\appendix

%%%%%%%%%%%%%%%%%%%%%%%%%%%%%%%%%%%%%%%%%%%%

\section{Methods}

\textbf{Preparation of domain-controlled heterostructure samples.} Domain-controlled samples were prepared by annealing an Fe-based amorphous metal, specifically Metglas (2605 SA1 for main data, Proterial), in an inert atmosphere at various temperatures. The amorphous metals were cut into pieces of approximately 3 mm $\times$ 6 mm and loaded into quartz ampoules. To prevent oxidation, the ampoules were sealed under a partial 6N Ar atmosphere. Two sets of samples were prepared with two different annealing times of 1 h and 5 min. The 1-h annealed samples were heated in a furnace at a ramp rate of 5 K/min and maintained for 1 h at temperatures of 673, 698, 723, 773, and 973 K, respectively. The 5-min annealed samples were heated in a furnace at a ramp rate of 10 K/min and maintained for 5 min at temperatures of 623, 673, 723, and 823 K, respectively. Finally, the 1-h annealed samples were naturally cooled in the furnace, while the 5-min annealed samples were water quenched.

\textbf{Characterization of the structural disorder using XRD.} The crystallinity and structural disorder of the amorphous samples were characterized using in-situ XRD (EMPYREAN, Marlvern Panalytical) with Cu K$\alpha$ radiation at 45 kV and 40 mA. The theta-2theta scan was performed after a 10-min hold at each temperature using a 255-channel strip detector (PIXcel$^{\text{1D}}$, Marlvern Panalytical) to minimize the dynamic change in crystallinity during the scan. The ramp rate was set at 5 K/min. The atmosphere for the in-situ measurements was filled with inert N$_{2}$ gas at a flow rate of 50 cc/min. Ex-situ XRD measurements (D/MAX-2500/PC, Rigaku) were conducted on pre-annealed samples at room temperature in the air with Cu K$\alpha$ at 40 kV and 100 mA.

\textbf{Imaging the microstructure and local disorder.} The microstructures of the samples were characterized using a scanning electron microscope (JSM-IT800HL, JEOL) combined with an EBSD detector (C-Nano, Oxford) at an acceleration voltage of 15 kV. Atomic-scale crystallinity was investigated using high-resolution TEM (JEM-2100F, JEOL) at an acceleration voltage of 200 kV. The TEM images were analyzed using Gatan software.

\textbf{Differential scanning calorimeter.} The exothermic reactions for recrystallization were investigated using DSC (DSC 200 F3 Maia, Netzsch) following the ASTM E-1269 standard. The temperature was increased from 300 K to 823 K at a heating rate of 5 K/min. The atmosphere was maintained at 5 N with N$_{2}$ gas during the measurements.

\textbf{Measurement of transport properties.} Electrical and thermal properties were measured using a cryogen-free measurement system (CFMS, Cryogenics). The electrical and Hall resistances were measured using an AC resistance bridge (Model 372, Lake Shore Cryogenics) with a current density of approximately 5 $\times$ 10$^{4}$ A/m$^{2}$. The thermal conductivity and Seebeck and Nernst effects were measured using an in-plane configuration incorporating two Peltier devices (NL 1010T-01AC, Marlow Industries) and two T-type thermocouples (Fig. S6 for details). A Peltier device was employed to pump heat into the sample using the Peltier effect with an electrical current of 100–150 mA supplied by a current source (E3646A, Agilent), whereas the other served as a heat flux sensor, measuring the heat flux using pre-calibrated heat flux sensitivity data for each temperature. Thermal grease (N \& H grease, Apiezon) and GE varnish (GE 7031, CMR direct) were used between the cryostat stage, Peltier devices, and samples to reduce thermal contact resistances and ensure uniform temperature distribution across the sample. The thermocouples were attached using silver epoxy (H20E, EPO-TEK) to measure the TE voltages (2182A, Keithley). All measurements utilized 25.4-diameter wires (SPCP- and SPCI-001-50, Omega Engineering) and were conducted under a high-vacuum condition ($<$10$^{-5}$ Torr) to minimize heat losses by conduction and convection.

\textbf{Measurement of magnetic properties.} Magnetic properties were measured using a SQUID-VSM (MPMS 3, Quantum Design) with a high-temperature oven option. The temperature was increased from 300 K to 760 K at a ramp rate of 5 K/min.

\textbf{Simulation of Seebeck-driven transverse thermoelectric generation (STTG).} The contribution of STTG to the enhanced $S_{yx}$ was investigated using a thermoelectric module in COMSOL Multiphysics software, combining Fourier’s law, Ohm’s law, and thermoelectric effects. A simplified composite model was introduced, assuming spherical crystalline structures embedded in an amorphous host (Fig. S14). The TT was considered by including off-diagonal terms (e.g., $\rho_{xy}$, and $S_{xy}$) in the conductivity and thermopower tensors~\cite{Bang24ApplEnergy}. The interfacial electrical and thermal contact resistances were neglected. 

\section{Supplemental Notes}

In the main text, we theoretically proposed the enhancement of transverse transport (TT) based on the physical mixture and showed experimental evidence of enhanced TT in a heterostructure. Here, in Supplemental Materials, we provide detailed theoretical and experimental information. Especially, we provide experimental evidence reinforcing the theoretical prediction that the enhanced TT originates from an extrinsic mechanism (i.e., the meandering path of electrons in the hetero-domain structures), not from intrinsic mechanisms. We do this by analyzing the ratio of anomalous Nernst conductivity ($\alpha_{yx}$) and anomalous Hall conductivity ($\sigma_{yx}$), and investigating possible contribution of the Seebeck-driven transverse thermoelectric generation (STTG) to the observed anomalous Nernst coefficient ($S_{yx}$). Additionally, the effect of domain control on longitudinal electrical transport, including the residual resistivity ratio (RRR), Kondo-like resistivity upturns, and magnetoresistance (MR), is analyzed and discussed. 

\subsection{Network size dependence of the transverse conductivity}

We evaluate the network size dependence of the transverse conductivity $\gamma_{yx}$, which is presented in Fig. 1(d) in the main text. The average $\bar{\gamma}_{yx}$ of $\gamma_{yx}$ over 10 random arrangements (instead of 100) is presented as a function of $P^{(B)}$ for each $N_{x}$ and $N_{y}$ in Fig. S1(a) to S1(d). The conductivity parameters are identical to those used in Fig. 1(d) in the main text. The values at $P^{(\text{B})}$ = 0\% or 100\% (without physical mixture) are identical regardless of the network size. These results are consistent with the fact that conductivity is an intrinsic property, which is independent of the system size. In the presence of the physical mixture, $\bar{\gamma}_{yx}$ (red symbols) is almost independent of the network size, but it slightly increases and show a larger deviation (red shaded area) as $N$ decreases. This is because the current paths and the arrangement of possible geometric structures of domains become restricted as the network size decreases. We note that the longitudinal conductivity $\gamma_{xx}$ also exhibits similar behavior to $\gamma_{yx}$.

\subsection{Investigation of structural transition upon annealing}

The domain distribution and crystallization behavior were first evaluated by in-situ X-ray diffraction (XRD) measurements on the as-cast sample during heating from 300 to 1073 K [Fig. S2(a)]. The onset of crystallization was identified at 723 K, above which distinct $\alpha$-Fe diffraction peaks appeared [Fig. S2(c)]. Further in-situ XRD analysis during cooling from 1073 K back to 300 K confirmed the persistence of these crystalline features, indicating the irreversible nature of the crystallization [Fig. S2(b)]. Minor diffraction peaks corresponding to Fe2B (PDF\#39-1314) and possible Fe-Si or Fe-B alloys phases were also observed at higher temperatures. We note that these changes in crystalline phases cannot explain the enhanced TT; a detailed discussion is provided in Note S6.

To further investigate the phase transition, differential scanning calorimetry (DSC) was conducted (Fig. S5), revealing two distinct exothermic peaks at $T_{x1}$=780 K and $T_{x2}$=820 K, characteristic of crystallization~\cite{Minor87JMS,McHenry03ScrMat}. The absence of transition peaks during the second DSC cycle also supports the irreversible nature of the observed exothermic reactions.

\subsection{Effect of domain control on longitudinal electrical transport}

First, we confirm that the controlled domain with varying annealing temperature ($T_{a}$) affects the longitudinal electrical transport. The electrical resistivity ($\rho_{xx}$) of the as-cast sample was 121 $\mu \Omega \cdot$cm at $T$ = 300 K. Intriguingly, annealing at $T_{a}$ = 673, 698 K increased $\rho_{xx}$ to 131–142 $\mu \Omega \cdot$cm [Fig. 4(a) in the main text and Fig. S8(a)], which can be attributed to the evolutions of local inhomogeneity and the (chemical) short-range order~\cite{Zhou15PNSM}. This implies that the degree of disorder that free carriers experience within the solid is more pronounced in heterostructure samples than in a fully amorphous solid (as-cast), which is consistent with observations in multi-element complex alloys~\cite{Tanimoto22JAC,Li23ActaMater,Rana03MCP}. Subsequently, $\rho_{xx}$ decreased to a saturated value ($\sim$60 $\mu \Omega \cdot$cm) in crystalline samples with high $T_{a}$ at 773 and 973 K.

The $\rho_{xx}$ for all the samples exhibited a typical $T$-dependence of metals, where $\rho_{xx}$ decreased with decreasing $T$ [Fig. S8(a)]. As electrons in metals mostly scatter with impurities at low temperatures rather than with phonons, RRR (= $\rho_{xx}$(300 K)/$\rho_{xx0}$, where $\rho_{xx0}$ is the residual resistivity owing to the static impurities at low temperatures, here at $T$=2 K) reflects the level of disorder in the samples. The amorphous samples exhibited a weak $T$ dependence on $\rho_{xx}$ with a small RRR of 1.04 compared with the crystallized $T_{a}$=973 K sample with an RRR of 2.08 [Fig. S10(c)]. This demonstrates the strong localization of electrons by the disorder in amorphous samples. We note that the RRR value of the $T_{a}$=973 K sample is still smaller than that of Fe single- and polycrystalline samples (RRR of $\sim$7–86)~\cite{Watzman16PRB}, indicating existing structural or chemical disorders, as observed in Fig. 3 in the main text. In the heterostructure samples, upturns in $\rho_{xx}$ were observed below 10–20 K, followed by a logarithmic $T$ dependence down to 2 K [Figs. S10(e) – S10(g)], indicative of unconventional scattering affecting mobile carriers at low temperatures. These mechanisms include the Kondo effect~\cite{Dobrosavljevic92PRL,Molinari23ACSAEM} owing to electron coupling with localized spin impurities, weak localization (WL)–induced quantum interference of electrons~\cite{Li21Matter,Lu15PRB}, or electron–electron interaction~\cite{Lu15PRB,Lee85RMP}, all of which indicate the presence of local disorders. In low temperature regime, WL may be responsible for weak conduction at low magnetic fields and its annealing temperature dependence. Notably, this upturn became negligible in the relatively more crystalline samples annealed above 723 K [Fig. S10(h) and S10(i)].

WL and weak antilocalization (WAL) describe the destructive and constructive quantum interferences of electron waves around self-intersecting paths, leading to increases and decreases in the $\rho_{xx}$, respectively~\cite{Molinari23ACSAEM,Li21Matter,Lee85RMP}. Under a magnetic field, these self-intersecting effects are mitigated, resulting in the negative MR for WL and positive MR for WAL. Therefore, analyzing MR can provide the information on how the carriers are affected by disorders within solids. The domain-controlled samples show a crossover of MR at a high field regime ($>$ 3 T) from negative (in as-cast sample) to positive MR ($T_{a}$=973 K) at $T$=2 K (Fig. S11), suggesting the transition from WL to WAL as $T_{a}$ increases, consistent with the $T$-dependent data presented above.

\subsection{Effect of domain control on thermal conductivity}

As ANE-based transverse thermoelectrics are attracting much attention nowadays~\cite{Uchida22Joule,Boona21JAP,Ikhlas17NatPhys,Sakai20Nature,Sakai18NatPhys,Pan22NatMater,He21Joule,Fu18EES,Guin19AdvMat}, we further evaluated the thermal conductivities ($\kappa$) of our samples to see their potential for thermoelectric applications. The $\kappa$ of as-cast and heterostructure samples at 300 K exhibits almost constant values around 11 W/(m$\cdot$K) and shows an abrupt rise to 22 W/(m$\cdot$K) when $T_{a}$ reaches 773 K (Fig. S9). Such a rapid transition in $\kappa$ can be explained by the amorphous-to-crystalline structural transition with increasing $T_{a}$, which is consistently observed in the lattice thermal conductivity ($\kappa_{\text{lat}}$) calculated from the Wiedemann-Franz law (Fig. S9). 

Our heterostructure samples demonstrate notably low $\kappa_{\text{lat}}$ of 3.3–5.8 W/(m$\cdot$K) ($T_{a}$= 673–723 K), contrary to single crystals known for substantial ANE, such as Co$_{2}$MnGa ($\kappa_{\text{lat}} \sim$17.5 W/(m$\cdot$K)~\cite{Guin19NPGAM}), Fe$_{3}$Ga ($\kappa_{\text{lat}} \sim$10.4 W/(m$\cdot$K)~\cite{Sakai20Nature}), and Fe-based Heusler compounds ($\kappa_{\text{lat}} \sim$9.4–10.6 W/(m$\cdot$K)~\cite{Mende21AdvSci}) at 300 K. This is attributed to the absence of long-range lattice order, resulting in localized vibrational modes~\cite{Zhou20AdvFuncMater}. 

As a result, the samples showing the enhanced ANE, specifically at $T_{a}$ = 698 K for $S_{yx}$ and $T_{a}$ = 723 K for $\alpha_{yx}$, also exhibited total thermal conductivity around 11 W/(m$\cdot$K), which is significantly lower than values reported for state-of-the art samples such as topological Weyl semimetals (e.g., Co$_{2}$MnGa: $\kappa$ $\sim$22 W/(m$\cdot$K)~\cite{Guin19NPGAM}, Fe-based binary alloys: $\sim$24 W/(m$\cdot$K) and Fe-based Heusler compounds: $\sim$18–20 W/(m$\cdot$K)~\cite{Mende21AdvSci}). 

Such relatively low $\kappa$ combined with competitive $S_{yx}$ values make the heterostructure samples advantageous for transverse thermoelectric energy conversion, where they can potentially achieve a high transverse figure-of-merit ($z_{\text{TT}} T = \frac{S_{yx}^{2} \sigma_{yy}}{\kappa} T$) for power generation or a high heat flux sensitivity (=$S_{yx}/\kappa$) for heat sensing applications~\cite{Zhou20APE}. 

We expect that incorporating heavy elements, such as Pt, could significantly reduce the thermal conductivity by enhancing phonon scattering through anharmonicity and localization of lattice vibrations. In addition, such heavy elements may increase spin-orbit coupling, thereby strengthening transverse scattering. These effects may offer pathways to optimize performance.

\subsection{Ratio of anomalous Nernst conductivity and anomalous Hall conductivity}

Next, we discuss the microscopic origin of ANE by analyzing the ratio of $\alpha_{yx}$ and $\sigma_{yx}$, as proposed by Xu et al.~\cite{Xu20PRB} According to the Berry curvature ($\Omega_{z}$) description on TT, $\sigma_{yx}$ and $\alpha_{yx}$ can be expressed as 

\begin{align}
    \sigma_{xy} &= \frac{e^{2}}{\hbar}\int_{\text{BZ}}\frac{d^{3}k}{(2\pi)^{3}}f(k) \Omega_{B}, \tag{S1}
    \\
    \alpha_{xy} &= \frac{ek_{B}}{\hbar}\int_{\text{BZ}}\frac{d^{3}k}{(2\pi)^{3}}s(k) \Omega_{B}, \tag{S2}
\end{align}

where $f(k)$ is the Fermi-Dirac distribution and $s(k) = -f(k) \, \text{ln} f(k) - (1-f(k)) \, \text{ln} (1 - f(k))$ is the entropy density of electron gas. Given the distinct distribution functions for AHE and ANE, $\sigma_{yx}$ is obtained by an integration of $\Omega_{B}$ for all the occupied bands below $E_{F}$, while $\alpha_{yx}$ is determined by $\Omega_{B}$ near $E_{F}$. Equations S1 and S2 can be rewritten~\cite{Xu20PRB,Ding19PRX} with proper length parameters of the Fermi wavelength $\lambda_{F}$ and the de Broglie thermal wavelength $\Lambda = \sqrt{\frac{h^{2}}{2\pi mk_{B}T}}$ : 

\begin{align}
    \sigma_{xy} &\approx \frac{e^{2}}{\hbar}\frac{1}{c} \left\langle \frac{\Omega_{B}}{\lambda^{2}_{F}} \right\rangle , \tag{S3}
    \\
    \alpha_{xy} &\approx \frac{ek_{B}}{\hbar}\frac{1}{c}\left\langle \frac{\Omega_{B}}{\Lambda^{2}} \right\rangle , \tag{S4}
\end{align}

where $m$ and $c$ are the mass of the particle and the lattice parameter along the applied magnetic field, respectively. In a high $T$ regime, where $\lambda_{F}$ becomes comparable with $\Lambda$, the $\alpha_{yx}/\sigma_{yx}$ ratio approaches $k_{B}/e=86$ $\mu$V/K. This tendency of $\alpha_{yx}/\sigma_{yx}$ near 300 K has well explained the TT in topological materials, whose large TT originate from the intrinsic Berry curvature, such as Co$_{2}$MnGa~\cite{Xu20PRB}. In contrast, YbMnBi$_{2}$ showed a large discrepancy of the ratio from $k_{B}/e$, which was explained by possible extrinsic contributions due to the large SOC~\cite{Pan22NatMater}. Despite the small SOC of our samples without heavy elements, amorphous and hetero-domain samples show discrepancies from $k_{B}/e$ near 300 K, indicating significant role of extrinsic mechanism (Fig. S12). By contrast, the crystalline sample ($T_{a}$=973 K) shows an increasing trend of $\alpha_{yx}/\sigma_{yx}$, indicating dominant intrinsic mechanism. In summary, the implication derived from this result is consistent with our other observations, including the resistivity upturn and crossover of MR, in that they all indicate a crucial role of extrinsic mechanism in our heterostructure samples. In addition, as previously discussed regarding the relation between $\rho_{yx}$ and $M$, neither large $S_{yx}$ nor $\alpha_{yx}$ are attributed to the change in $M$ (Fig. S13), making a distinction from conventional ANE materials whose $S_{yx}$ scales with $M$~\cite{Guin19NPGAM}.

\subsection{Contribution of the Seebeck-driven transverse voltage}

Lastly, we briefly discuss other possibility that may increase $S_{yx}$. The hetero-domain samples can be considered as a composite of amorphous and crystalline phases. In such a composite structure, the longitudinal Seebeck voltage in one material (e.g., crystal) can contribute ANE through AHE in the other material (e.g., amorphous), a phenomenon referred to as the Seebeck-driven thermoelectric generation (STTG)~\cite{Zhou21NatMater}. Given the substantial anomalous Hall angle ($\sim$6.5 \%) of the amorphous material (as-cast), the large $S_{yx}$ would be attributed to STTG if the Seebeck coefficient ($S_{xx}$) of crystalline materials is sufficiently large. To quantify the contribution of STTG on enhanced $S_{yx}$, we perform a COMSOL simulation using a simplified composite model assuming spherical composites (Fig. S14 and Methods). Given small amount of B (2.5 at.\%), we considered stable Fe- and Si-based crystalline phases, whose material properties were taken from literatures~\cite{Zhou20APE,Secco17PEPI,Hsin17Nanotech,Buschinger97PhysicaB,Hamada21APL}. Figure S14 shows that the STTG effect in our samples is small and even decreases the total $S_{yx}$ when the amorphous host is mixed with Fe, Fe$_{3}$Si, Fe$_{2}$Si, and FeSi, possibly due to the opposite sign in $S_{yx}$ or electrical shunting. When FeSi$_{2}$ forms the composite with the amorphous host, $S_{yx}$ increase, but the effect is negligible (less than 0.03 $\mu$V/K at the volume ratio of crystalline material ($V_{\text{c}}$) of 7\%. Note that considering the small atomic portion of Si (5 at.\%), the $V_{\text{c}}$ of 7\% is the maximum that FeSi$_{2}$ can form in the host. Thus, we conclude that the contribution of STTG to the observed enhanced TT in our heterostructure samples samples is negligibly small.

\section{Supplemental Figures}

%%%%%%%%%%%%%%%%<Figure>%%%%%%%%%%%%%%%%
\begin{figure}[h]
\includegraphics[width=400pt]{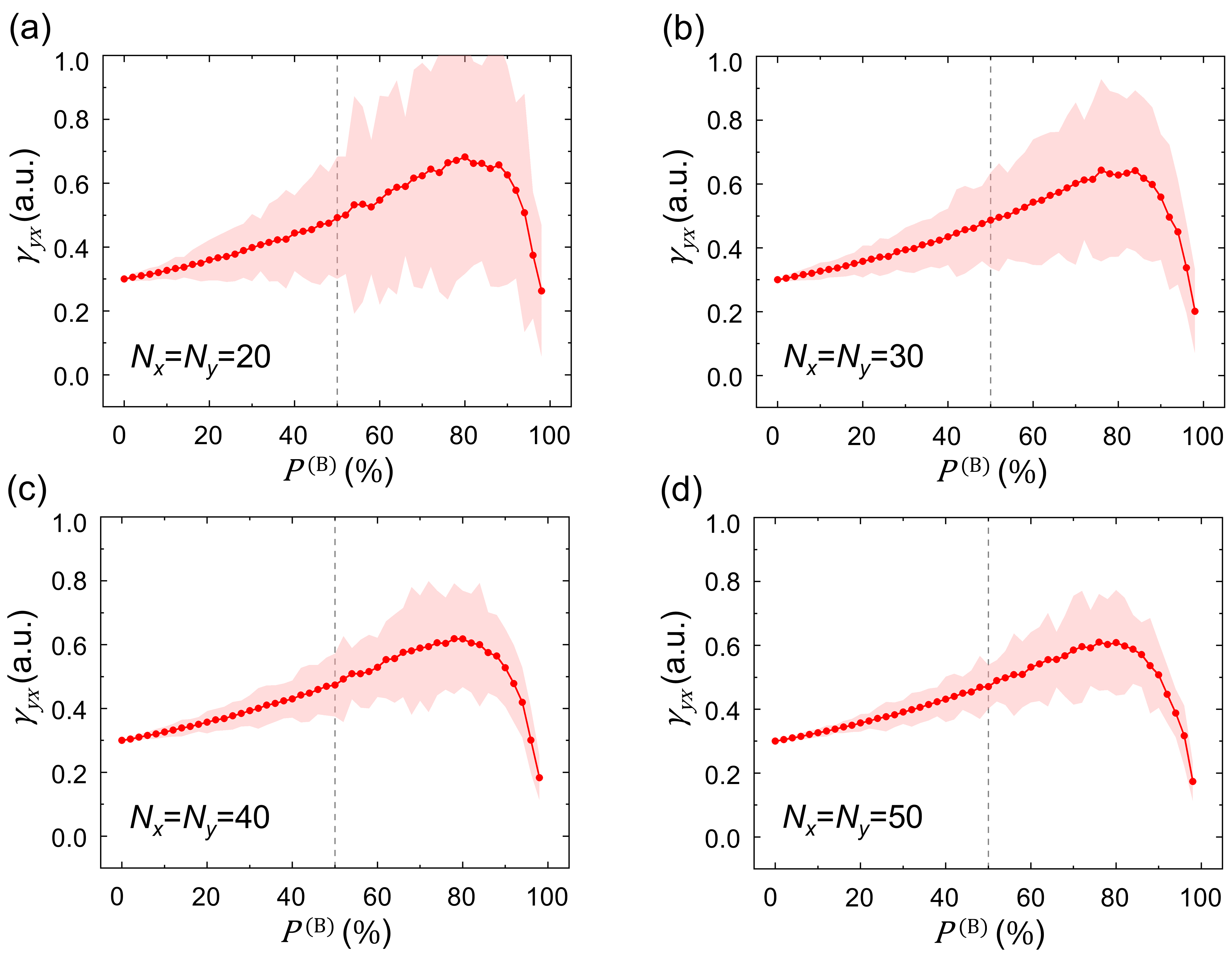}
\caption{\label{figS1}Network size dependence of the transverse conductivity. $\bar{\gamma}_{yx}$ (red symbols) and its deviation (red shaded area) are presented for the network size of $N_{x} = N_{y}$ = (a) 20, (b) 30, (c) 40, and (d) 50.}
\end{figure}
%%%%%%%%%%%%%%%%%%%%%%%%%%%%%%%%%%%%%%%%

\clearpage

%%%%%%%%%%%%%%%%<Figure>%%%%%%%%%%%%%%%%
\begin{figure}[h]
\includegraphics[width=400pt]{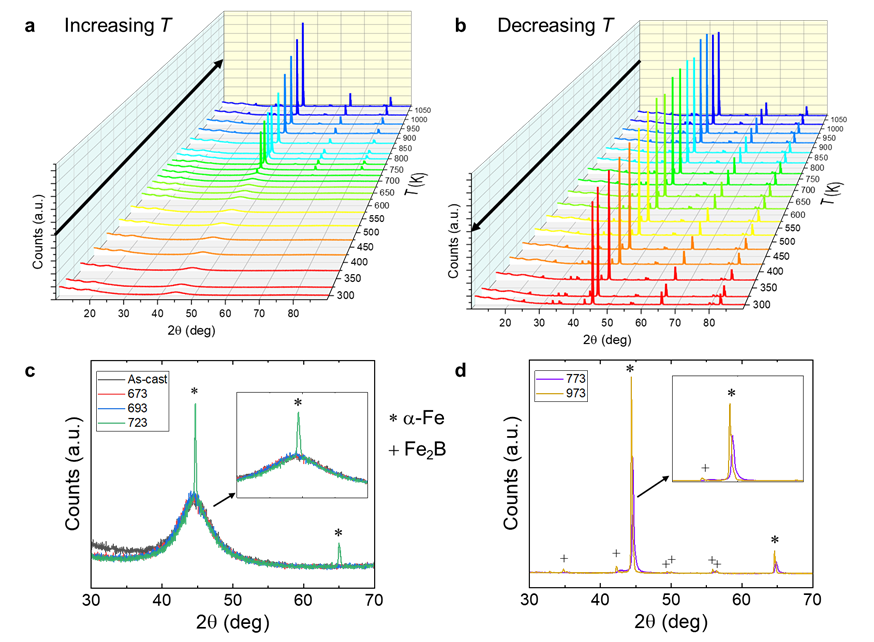}
\caption{\label{figS2}In-situ X-ray diffraction patterns of the as-cast sample. In-situ X-ray diffraction patterns for (a) increasing temperature from 300 K to 1073 K and (b) decreasing from 1073 K to 300 K. (c),(d) XRD patterns for main samples presented in the main text. }
\end{figure}
%%%%%%%%%%%%%%%%%%%%%%%%%%%%%%%%%%%%%%%%

\clearpage

%%%%%%%%%%%%%%%%<Figure>%%%%%%%%%%%%%%%%
\begin{figure}[h]
\includegraphics[width=400pt]{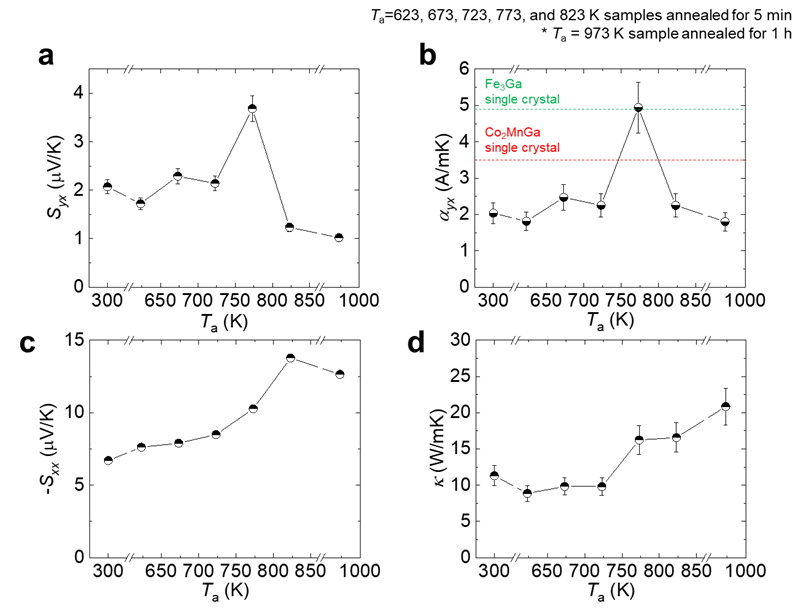}
\caption{\label{figS3}Thermal transport properties in the sample set annealed for 5 min. (a) Anomalous Nernst coefficient ($S_{yx}$) and (b), anomalous Nernst conductivity ($\alpha_{yx}$). The horizontal lines in (B) correspond to the values of Fe$_{3}$Ga and Co$_{2}$MnGa single crystals. (c) Longitudinal Seebeck coefficient with negative sign ($-S_{xx}$). (d) Longitudinal thermal conductivity ($\kappa$). All properties were measured at $T$=300 K. Note that data of the $T_{a}$ = 973 K correspond to the sample annealed at $T_{a}$ = 973 K for 1 h, served as the crystalline counterpart for comparison. The other data correspond to the samples annealed for 5 min aside from the as-cast sample.}
\end{figure}
%%%%%%%%%%%%%%%%%%%%%%%%%%%%%%%%%%%%%%%%

\clearpage

%%%%%%%%%%%%%%%%<Figure>%%%%%%%%%%%%%%%%
\begin{figure}[h]
\includegraphics[width=400pt]{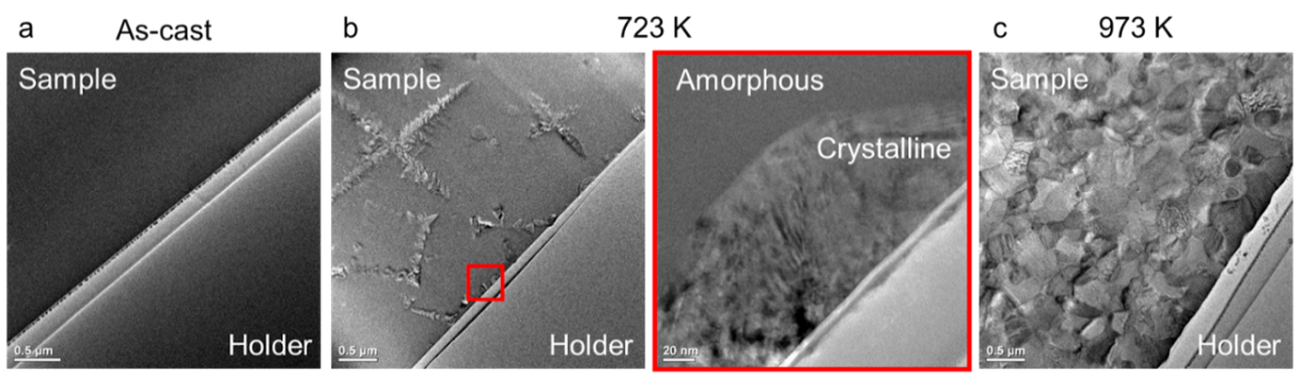}
\caption{\label{figS4}Low magnification images of TEM samples. Images of (a), the as-cast sample, and samples annealed at (b), 723 K (hetero-domain) and (c), 973 K (fully crystallized).}
\end{figure}
%%%%%%%%%%%%%%%%%%%%%%%%%%%%%%%%%%%%%%%%

\clearpage

%%%%%%%%%%%%%%%%<Figure>%%%%%%%%%%%%%%%%
\begin{figure}[h]
\includegraphics[width=300pt]{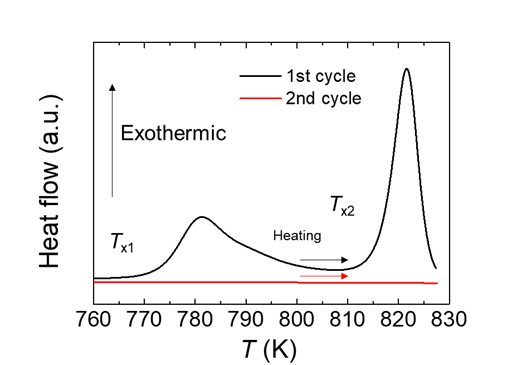}
\caption{\label{figS5}Differential scanning calorimetry with two-step exothermic reactions for crystallizations at $T_{x1}$ and $T_{x2}$. No peak is observed in the consecutive second cycle, confirming the irreversibility of the reactions. Both cycles were conducted while increasing the temperature at a rate of 5 K/min (Methods).}
\end{figure}
%%%%%%%%%%%%%%%%%%%%%%%%%%%%%%%%%%%%%%%%

\clearpage

%%%%%%%%%%%%%%%%<Figure>%%%%%%%%%%%%%%%%
\begin{figure}[h]
\includegraphics[width=400pt]{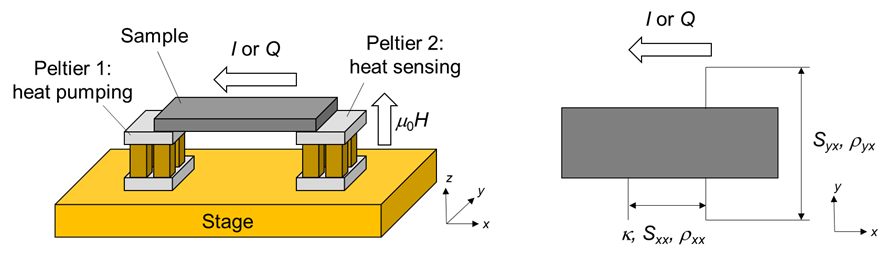}
\caption{\label{figS6}Measurement configurations for electrical and thermal transport properties.}
\end{figure}
%%%%%%%%%%%%%%%%%%%%%%%%%%%%%%%%%%%%%%%%

\clearpage

%%%%%%%%%%%%%%%%<Figure>%%%%%%%%%%%%%%%%
\begin{figure}[h]
\includegraphics[width=500pt]{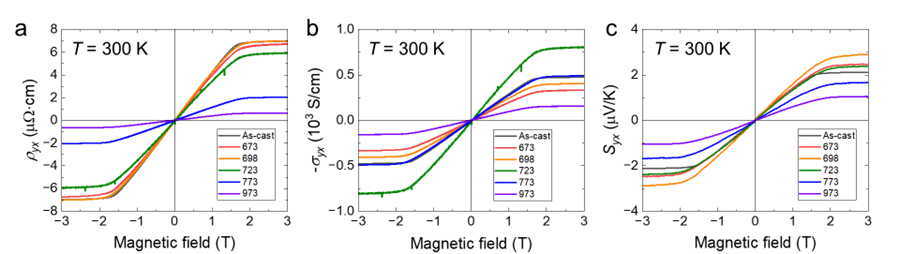}
\caption{\label{figS7}Magnetic field dependence of transverse transport properties, corresponding to Fig. 3 in the main text. (a) Anomalous Hall resistivity ($\rho_{yx}$), (b) anomalous Hall conductivity shown as $-\sigma_{yx}$, and (c) anomalous Nernst coefficient ($S_{yx}$) of the samples measured at T = 300 K. The property values in Fig. 3 were extracted from field-saturated regime beyond $\pm2.5$ T after eliminating the ordinary Hall and Nernst components by linear fittings.}
\end{figure}
%%%%%%%%%%%%%%%%%%%%%%%%%%%%%%%%%%%%%%%%

\clearpage

%%%%%%%%%%%%%%%%<Figure>%%%%%%%%%%%%%%%%
\begin{figure}[h]
\includegraphics[width=500pt]{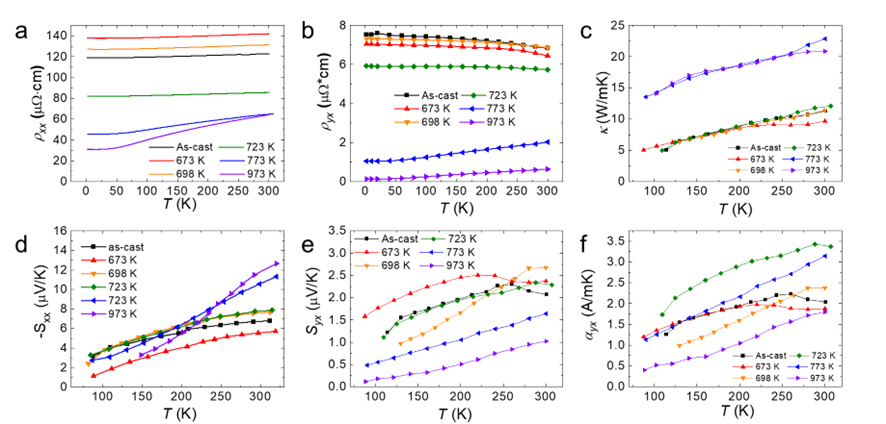}
\caption{\label{figS8}Temperature dependent longitudinal and transverse transport properties. (a) electrical resistivity ($\rho_{xx}$), (b) anomalous Hall resistivity ($\rho_{yx}$), (c) thermal conductivity ($\kappa$), (d) Seebeck coefficient with negative sign ($-S_{xx}$), (e) anomalous Nernst coefficient ($S_{yx}$), and (f) anomalous Nernst conductivity ($\alpha_{yx}$).}
\end{figure}
%%%%%%%%%%%%%%%%%%%%%%%%%%%%%%%%%%%%%%%%

\clearpage

%%%%%%%%%%%%%%%%<Figure>%%%%%%%%%%%%%%%%
\begin{figure}[h]
\includegraphics[width=350pt]{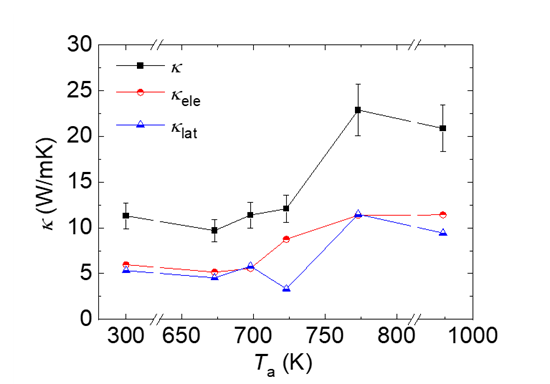}
\caption{\label{figS9}Total ($\kappa$), electronic ($\kappa_{\text{ele}}$), and lattice ($\kappa_{\text{lat}}$) thermal conductivities of 1-hr annealed samples at room temperature.  }
\end{figure}
%%%%%%%%%%%%%%%%%%%%%%%%%%%%%%%%%%%%%%%%

\clearpage

%%%%%%%%%%%%%%%%<Figure>%%%%%%%%%%%%%%%%
\begin{figure}[h]
\includegraphics[width=500pt]{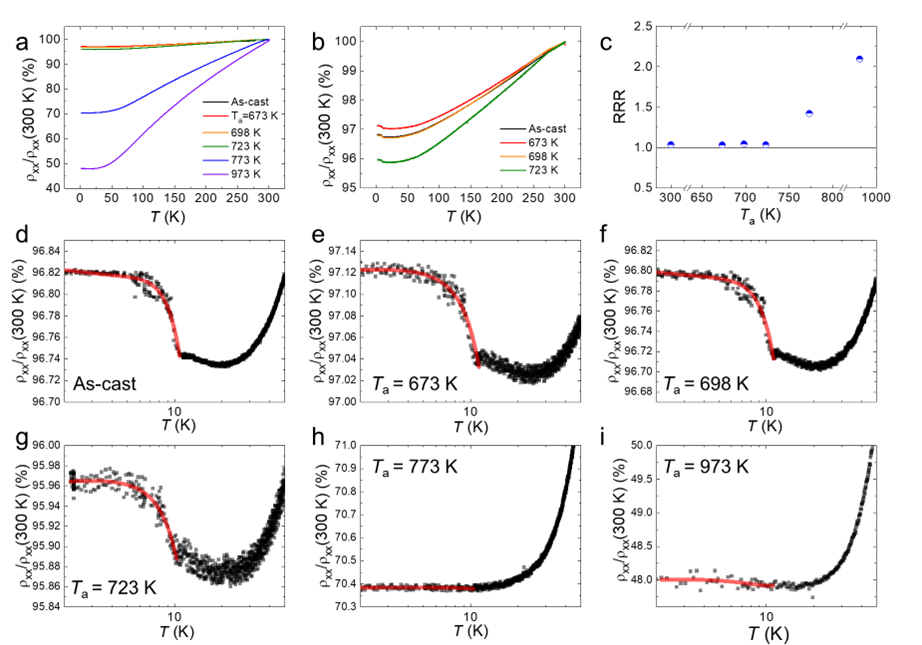}
\caption{\label{figS10}Temperature dependent longitudinal electrical properties. (a and b) Electrical resistivity ($\rho_{xx}$) normalized by 300 K data $\rho_{xx}$(300K) for all samples (a) and amorphous and hetero-domain samples (b). (c) Residual resistivity ratio (RRR= $\rho_{xx}$(300 K)/$\rho_{xx}$(2 K)). (d to i) $\rho_{xx}$/$\rho_{xx}$(300 K) at low temperature from 2 K to 50 K in as-cast (d), samples annealed at  673 K (e), 698 K (f), 723 K (g), 773 K (h), and 973 K (i). The red lines are the log functions for fitting. The resistivity upturns in amorphous and hetero-domain samples indicate the Kondo-like effects, including the Kondo effect, electron-electron interaction, and weak (anti)localization.}
\end{figure}
%%%%%%%%%%%%%%%%%%%%%%%%%%%%%%%%%%%%%%%%

\clearpage

%%%%%%%%%%%%%%%%<Figure>%%%%%%%%%%%%%%%%
\begin{figure}[h]
\includegraphics[width=350pt]{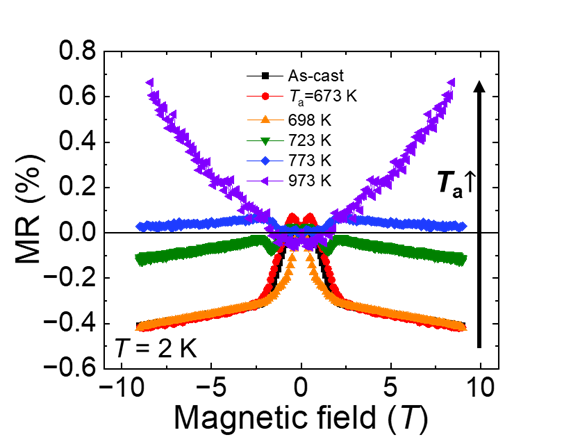}
\caption{\label{figS11}Magnetoresistance (MR) at T = 2 K. The MR data show a transition from negative MR to positive MR with increasing $T_{a}$, which indicates a crossover of quantum interference of electron waves from weak localization to weak antilocalization.}
\end{figure}
%%%%%%%%%%%%%%%%%%%%%%%%%%%%%%%%%%%%%%%%

\clearpage

%%%%%%%%%%%%%%%%<Figure>%%%%%%%%%%%%%%%%
\begin{figure}[h]
\includegraphics[width=350pt]{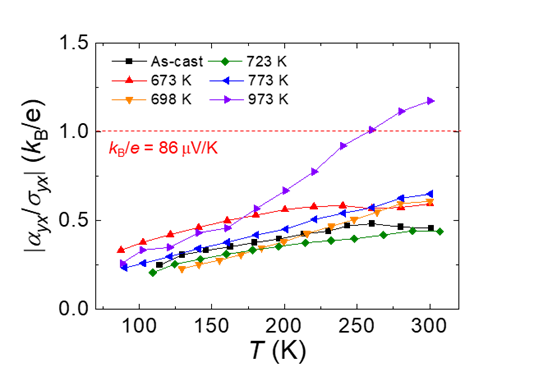}
\caption{\label{figS12}Ratio of anomalous Nernst conductivity and anomalous Hall conductivity.}
\end{figure}
%%%%%%%%%%%%%%%%%%%%%%%%%%%%%%%%%%%%%%%%

\clearpage

%%%%%%%%%%%%%%%%<Figure>%%%%%%%%%%%%%%%%
\begin{figure}[h]
\includegraphics[width=500pt]{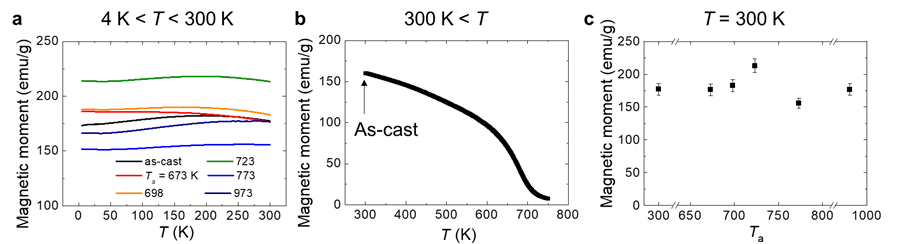}
\caption{\label{figS13}Magnetic properties measured by VSM. Temperature dependent magnetic moments of the samples ranging (a), from 4 K to 300 K, and (b), from 300 K to 750 K. Note that the crystallinity of as-cast sample changes during the high-temperature measurement in (a). (c) Magnetic moments of the samples at 300 K.}
\end{figure}
%%%%%%%%%%%%%%%%%%%%%%%%%%%%%%%%%%%%%%%%

\clearpage

%%%%%%%%%%%%%%%%<Figure>%%%%%%%%%%%%%%%%
\begin{figure}[h]
\includegraphics[width=400pt]{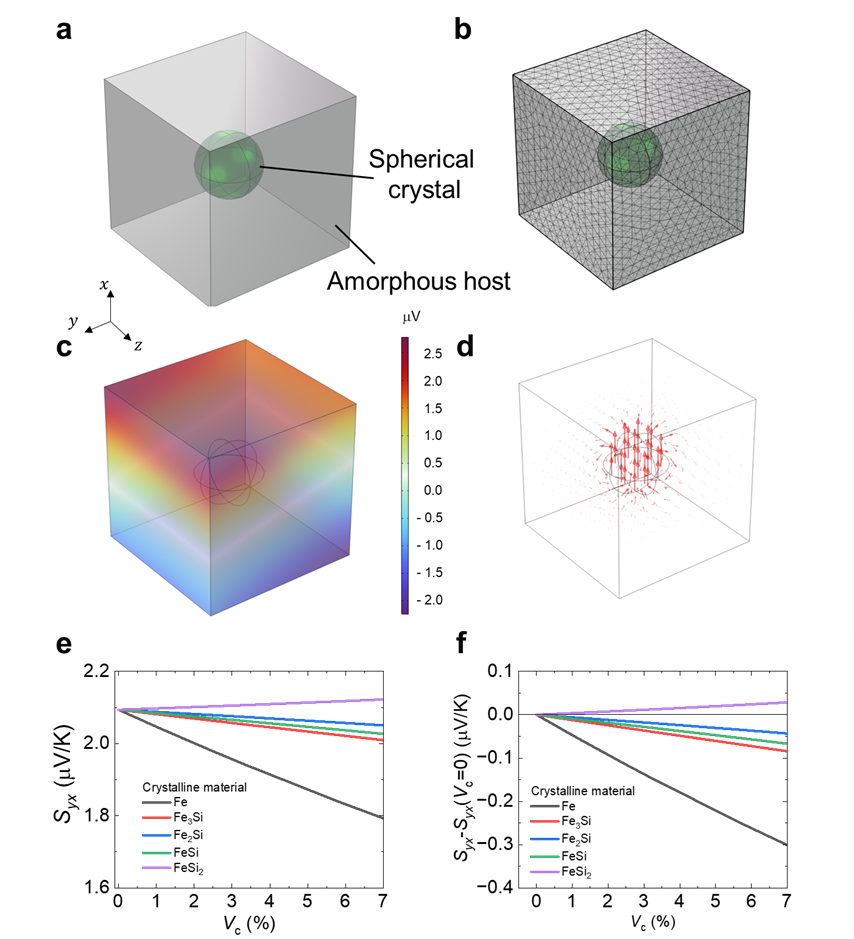}
\caption{\label{figS14}Investigation of the possible contribution of Seebeck-driven transverse thermoelectric generation (STTG). (a) Simplified composite model used in this simulation, assuming spherical crystal in amorphous host. (b) Constructed mesh for finite element method. (c and d) Simulated voltage distribution in the composite. (e) Anomalous Nernst coefficient ($S_{yx}$) considering the STTG effect as a function of crystalline volume ($V_{\text{c}}$) (f) Changes in $S_{yx}$ of composite compared to the as-cast sample ($S_{yx}$($V_{\text{c}}=0$)).}
\end{figure}
%%%%%%%%%%%%%%%%%%%%%%%%%%%%%%%%%%%%%%%%

\end{document}